\documentclass[12pt]{iopart}

\usepackage{color}
\usepackage{graphicx}
\usepackage{dcolumn}
\usepackage{bm}
\usepackage{psfrag}
\usepackage{epstopdf}
\usepackage{subfigure}
\usepackage{textcomp}
\usepackage{hyperref}
\usepackage{comment}
\usepackage{multirow}
\usepackage{amstext} 

\usepackage{iopams}

\newcommand{\dens}{$n_e$}

\newcommand{\nre}{$n_{\text{RE}}$}
\newcommand{\nAr}{$n_{\text{Ar}}$}
\newcommand{\densunit}{$\times 10^{19}$ m$^{-3}$}
\newcommand{\nArunit}{$\times 10^{20}$ m$^{-3}$}

\newcommand{\Ereavg}{$\left< E_{RE} \right>$}
\newcommand{\Teavg}{$\left< T_e \right>$}
\newcommand{\Te}{$T_e$}
\newcommand{\Ere}{$E_{RE}$}
\newcommand{\Egammax}{$E_{\gamma,max}$}

\newcommand{\Ip}{$I_P$}
\newcommand{\Ire}{$I_{RE}$}

\newcommand{\dB}{$\delta B$}

\newcommand{\Jpar}{$J_{||}$}
\newcommand{\Vloop}{$V_{\text{loop}}$}

\begin{document}



\title[]{Runaway Electron Seed Formation at Reactor-Relevant Temperature}

\author{C. Paz-Soldan,$^1$ P. Aleynikov,$^2$, E. M. Hollmann,$^3$ A. Lvovskiy,$^1$  I. Bykov$^3$, X. Du,$^1$, N. W. Eidietis,$^1$  D. Shiraki$^4$}

\address{$^1$General Atomics, San Diego, CA 92186-5608, USA}
\address{$^2$Max-Planck Institute for Plasma Physics, Greifswald, Germany}
\address{$^3$University of California San Diego, La Jolla, CA 92093-0417, USA}
\address{$^4$Oak Ridge National Laboratory, Oak Ridge, TN 37831, USA}

\date{\today}

\begin{abstract}
Systematic variation of the pre-disruption core electron temperature (\Te{}) from 1 to 12 keV using an internal transport barrier scenario reveals a dramatic increase in the production of `seed' runaway electrons (REs), ultimately accessing near-complete conversion of the pre-disruption current into sub-MeV RE current. Injected Ar pellets are observed to ablate more intensely and promptly as \Te{} rises. At high \Te{}, the observed ablation exceeds predictions from published thermal ablation models. Simultaneously, the thermal quench (TQ) is observed to significantly shorten with increasing \Te{} - a surprising result.  While the reason for the shorter TQ is not yet understood, candidate mechanisms include: insufficiently accurate thermal ablation models, enhanced ablation driven by the seed RE population, or significant parallel heat transport along stochastic fields. Kinetic modeling that self-consistently treats the plasma cooling via radiation, the induced electric field, and the formation of the seed RE is performed. Including the combined effect of the inherent dependence of hot-tail RE seeding on \Te{} together with the shortened TQ, modeling recovers the progression towards near-complete conversion of the pre-disruption current to RE current as \Te{} rises.  Measurement of the HXR spectrum during the early current quench (CQ) reveals a trend of decreasing energy with pre-disruption \Te{}. At the very highest \Te{} ($\approx$ 12 keV), $\approx$ 100\% conversion of the thermal current to runaway current is found. The energy of this peculiar RE beam is inferred to be sub-MeV as it emits vanishingly few MeV hard X-rays (HXRs).  These measurements demonstrate novel TQ dynamics as \Te{} is varied and illustrate the limitations of treating the RE seed formation problem without considering the inter-related dependencies of the pellet ablation, radiative energy loss, and resultant variations of the TQ duration. If the observed shortening of the TQ with increasing \Te{} extends to fusion-grade plasmas, than their propensity to form large quantities of RE seeds at high \Te{} may be far worse than previously thought. Positively, the high \Te{} scenario in DIII-D produces REs so prodigiously that it can serve as a meaningful new platform for demonstrating RE avoidance techniques.
\end{abstract}



\maketitle



\section{Introduction and Motivation}
\label{sec:intro}

The production of relativistic `runaway' electrons (RE) during a tokamak disruption is a grave concern for future fusion-grade tokamaks such as ITER \cite{Hender2007,Lehnen2014,Hollmann2015,Boozer2015,Breizman2017,Breizman2019}. Central to this concern is the expected large avalanche multiplication factor of any RE `seed', which scales exponentially with plasma current (\Ip{}) \cite{Rosenbluth1997,Boozer2018} and may be enhanced by interactions with high-Z impurities \cite{Martin-Solis2015,Hesslow2019}. A robust program of experimentation on existing tokamaks is presently ongoing to tackle this challenge \cite{Papp2013,Hollmann2013,Granetz2014,Reux2015,PazSoldan2017,Esposito2017,Zeng2017,Shevelev2018,
Mlynar2019,Carnevale2019,PazSoldan2019b,Ficker2019}.

When considering the extrapolation of RE dynamics observed on present devices to fusion-grade plasmas, a key sensitivity expected from first principles is the pre-disruption electron temperature (\Te{}). To ensure a strong cross-section for D-T fusion, and assuming equipartition of ion and electron energy, \Te{} is strongly constrained to be $\approx$ 10 keV or higher in a fusion-grade plasma \cite{Lawson1957}. In contrast, owing to various constraints present-day RE experiments are generally conducted in few-keV \Te{} plasmas or colder (the AUG tokamak being a notable exception \cite{Papp2016}). Despite the strong expected sensitivity of the RE dynamics on \Te{}, no dedicated experimental study has yet appeared in the literature, thus motivating this exploratory work.

\begin{figure}
\centering
\includegraphics[width=0.4\textwidth]{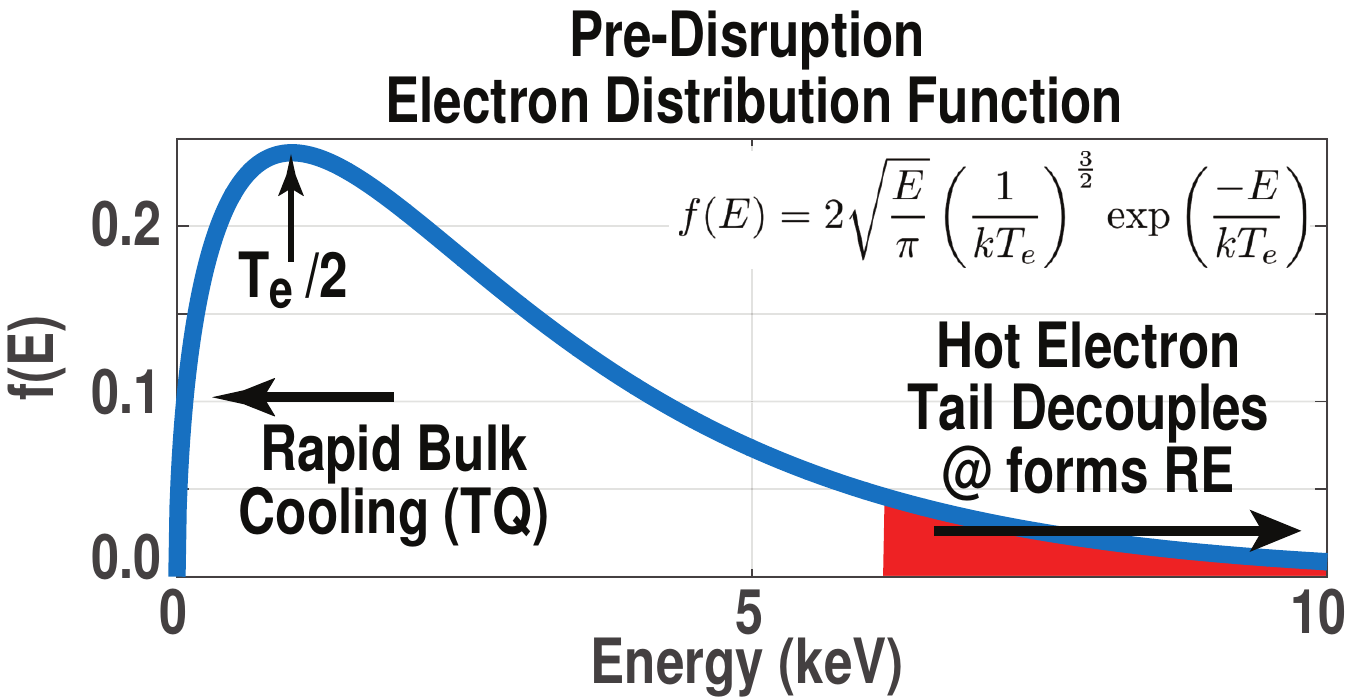}
\vspace{-10 pt}
\caption{Conceptual picture of the `hot-tail' mechanism.}
\label{fig:hottail}
\end{figure}

The strong expected sensitivity to \Te{} arises from the `hot-tail' mechanism \cite{Chiu1998,Smith2008a,Harvey2000,Helander2004,Smith2005}, illustrated conceptually in Fig. \ref{fig:hottail}. Hot-tail is expected to dominate fusion-grade plasma RE seed formation \cite{Smith2008a,Martin-Solis2017}. In this mechanism, the pre-disruption electron distribution contains a `tail' of high energy electrons that experience weaker collisional coupling to the bulk (due to the decreasing dependence of collisional drag on the thermal velocity in a plasma). During the rapid bulk cooling of the `Thermal Quench' (TQ), this tail population can decouple from the bulk if the rate of cooling exceeds the collisional equilibration time ($\propto T_e^{-\frac{3}{2}}$). Intuitively, the two key parameters governing in this mechanism are the pre-disruption \Te{} (which sets the size of the tail) and the TQ duration (which competes with the tail dissipation time).

Theory of this process was first presented in Ref. \cite{Smith2008a}, developed in the limit that the TQ duration is faster than the collisional equilibration time. This model thus effectively counts the number of electrons (red region in Fig. \ref{fig:hottail}) expected to be collisionally decoupled for an input TQ duration, which is pre-defined and not self-consistent with radiation cooling or transport. Nonetheless, this model is straightforward and commonly used to evaluate the `hot-tail' seed \cite{Smith2008a,Papp2013,Feher2011,Martin-Solis2017} A more recent model, presented in Ref. \cite{Aleynikov2017}, self-consistently calculates the TQ cooling dynamics consistent with the expected energy loss due to radiation. Both of these models are locally evaluated and contain no spatial transport effects. Going beyond these analytic models requires sophisticated 3D MHD computation to treat the transport, such as with NIMROD\cite{Izzo2011,Izzo2012}, M3D-C1\cite{Lyons2019}, and JOREK\cite{Bandaru2019,Sommariva2018}. Self-consistent treatment also formally requires interfacing the MHD timestep with a kinetic treatment of the non-Maxwellian RE generation and subsequent back-reaction on the Ampere-Faraday laws to capture possible interactions with the disruption MHD dynamics itself. Such a fully integrated treatment is outside of present capability though work is ongoing in this direction \cite{Hirvijoki2018,Harvey2019}. 

Thus far theoretical studies are generically characterized by an absence of direct experimental input, in particular regarding the interdependencies of the TQ cooling dynamic with other related effects such as the pellet ablation and the MHD dynamic. Providing the necessary input on what happens in experiment as \Te{} is systematically varied is a key goal of this study. For example: Does the RE seed generation change with \Te{}?  Does the pellet ablation increase enough with \Te{} to affect the cooling rate and thus the TQ duration? Does the kinetic model of Ref. \cite{Aleynikov2017} recover the experimental results, and using what approximations? 

\begin{figure}
\centering
\includegraphics[width=0.47\textwidth]{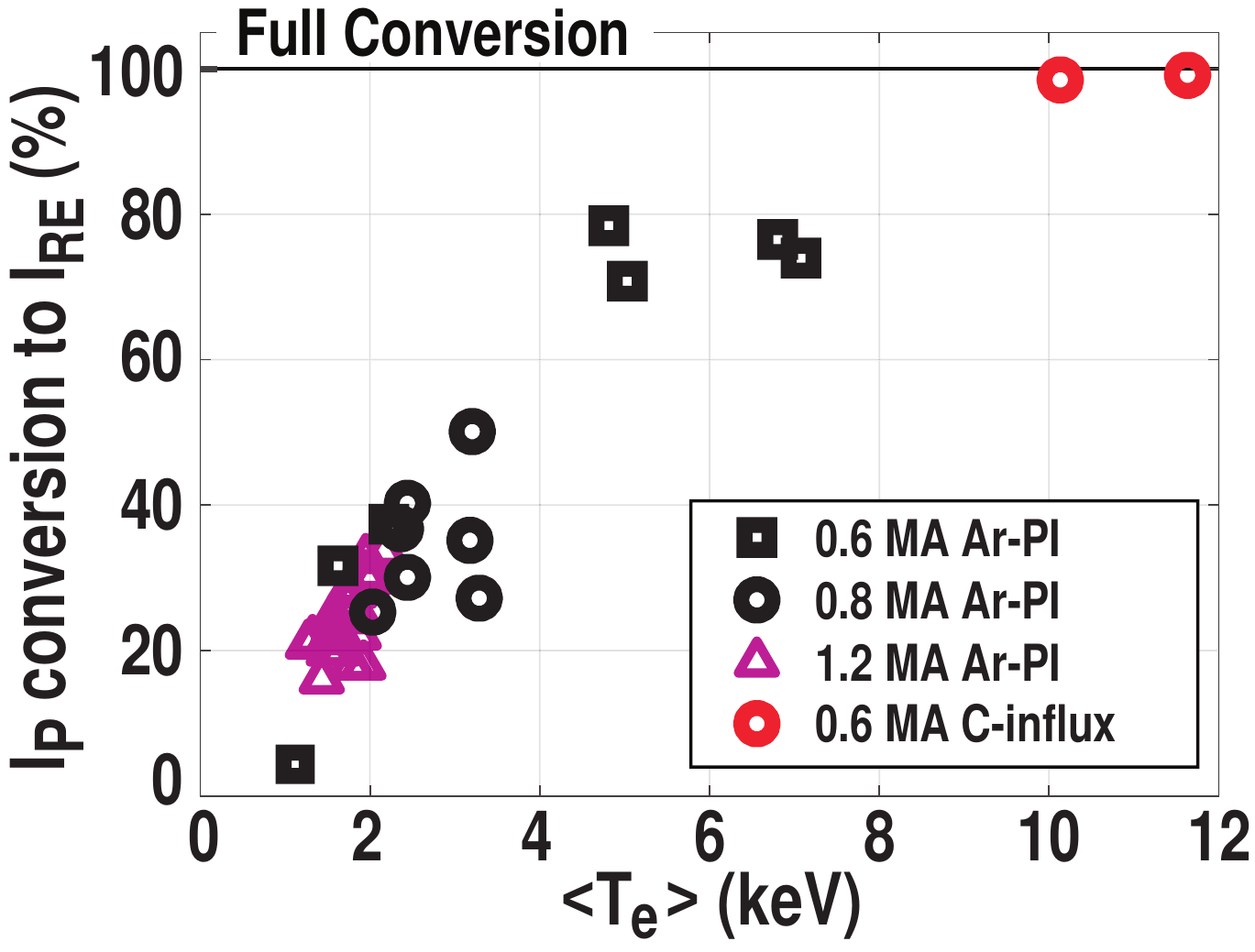}
\vspace{-10 pt}
\caption{A strong dependence of the conversion of pre-disruption current (\Ip{}) into RE current (\Ire{}) on pre-disruption \Te{} is observed. Explanation of this effect and its implications is the goal of this work.}
\label{fig:ipfig}
\end{figure}

In this paper, the experimental dynamics of RE seed formation is reported as \Te{} is systematically varied from 1 to 12 keV. Such high \Te{} is achieved by using an internal transport barrier scenario, and REs were formed using Ar pellet injection (Ar-PI). As shown in Fig. \ref{fig:ipfig}, a dramatic increase in the RE seed production with \Te{} is observed, culminating in full conversion of the pre-disruption current into RE current at the highest \Te{} (resulting in a sub-MeV RE beam). Reference discharges where RE were formed with Ar massive gas injection (MGI) showed no \Te{} effect in the explored range from 1-4 keV. With Ar-PI, further important TQ dynamics are revealed, namely an intense enhancement of the injected Ar pellet ablation rate, and a concomitant shortening of the TQ duration. These effects further support strong RE production as \Te{} is increased. Kinetic modeling of the `hot-tail' RE seed production in these discharges using the model of Ref. \cite{Aleynikov2017} is conducted. The main experimental trend of strongly increasing RE production with \Te{} is captured by the kinetic model when the effects of increased Ar ablation and shortened TQ duration are included. A non-monotonic dependence of the seed production on \Te{}, predicted by kinetic modeling only when Ar quantity is held fixed \cite{Aleynikov2017}, is not observed.

This work builds on the study of Ref. \cite{Hollmann2017} (summarized in Ref. \cite{PazSoldan2019b}), where the first estimate of the RE seed current and its comparison to modeling predictions was presented in a low \Te{} plasma. A RE seed current measurement was extracted and found to be over-predicted by the Smith model(Ref. \cite{Smith2008a}), yet under-predicted by the Aleynikov and Breizman model (Ref.\cite{Aleynikov2017}). Both models exhibited an exponential sensitivity to input parameters in the \Te{} range studied. This work extends the study of Ref. \cite{Hollmann2017} by examining systematically the effect of pre-disruption \Te{} on the pellet ablation, TQ duration, and RE seed formation processes, with consistent results found at low \Te{}.

The remainder of this paper is structured as follows. The experimental setup is described in Sec \ref{sec:setup}. The method to extract the RE seed current is described in Sec. \ref{sec:seed}. The observed Ar pellet ablation and thermal quench dynamics are described in Sec. \ref{sec:dynamics}. Kinetic modeling is described and compared to experiment in Sec. \ref{sec:model}.  Finally, full conversion dynamics at \Te{} $\approx$ 12 keV are described in Sec. \ref{sec:prompt}. Discussions and conclusions are given in Sec. \ref{sec:disc}. \ref{sec:gri} presents HXR spectra measured during the early current quench that show a decreasing RE energy with pre-disruption \Te{}.


\section{Experimental Setup}
\label{sec:setup}

\begin{figure}
\centering
\includegraphics[width=0.4\textwidth]{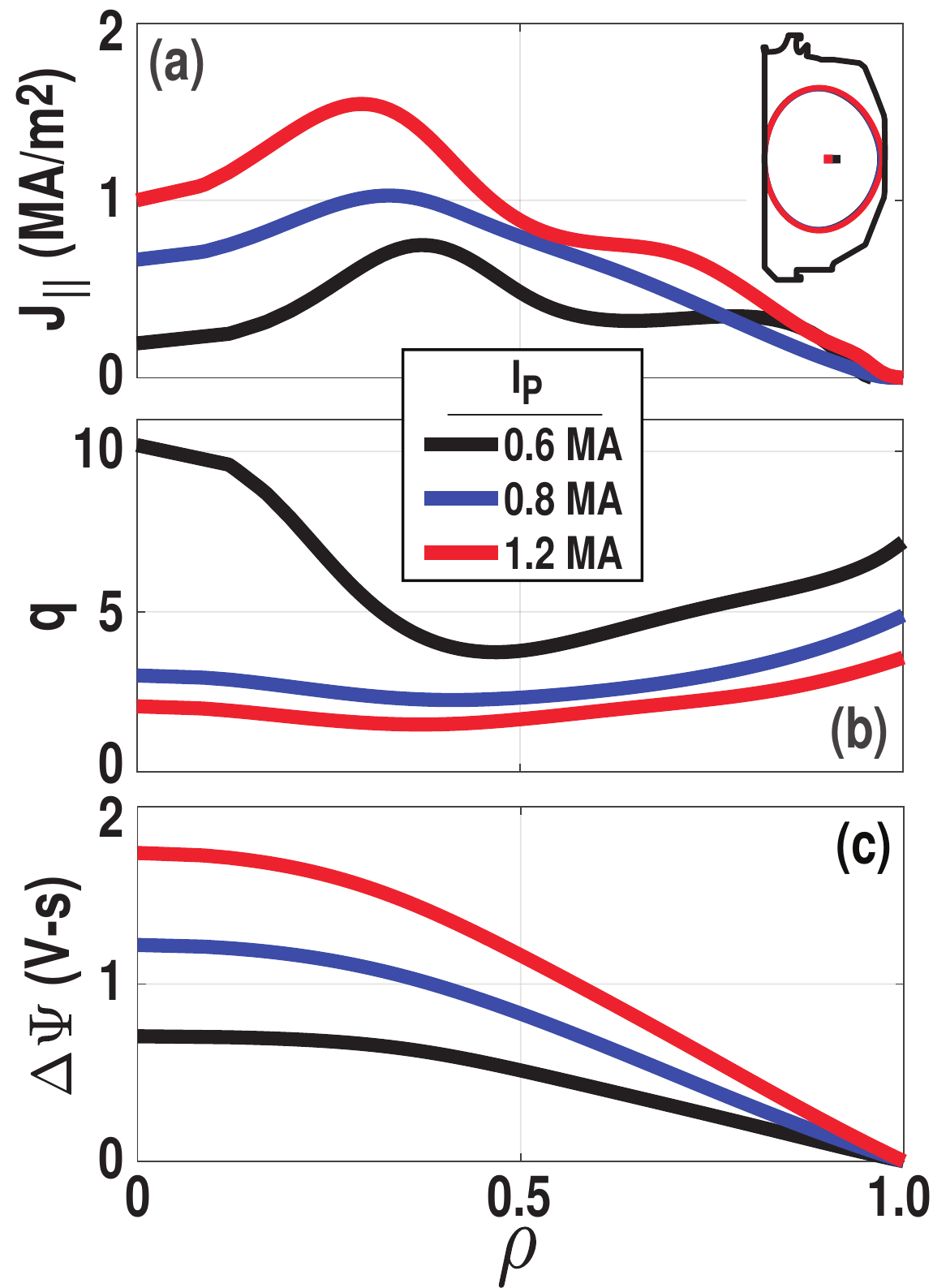}
\vspace{-10 pt}
\caption{Motional Stark Effect (MSE) constrained equilibrium reconstructions of (a) toroidal current (\Jpar{}) profile (b) safety factor (q) profile, and (c) poloidal flux change from the edge.}
\label{fig:EQs}
\end{figure}

Experiments are conducted in a variant of the conventional RE producing discharge on DIII-D \cite{Hollmann2017,Shiraki2018}. The discharge is an inner-wall limited low elongation plasma (see Inset of Fig. \ref{fig:EQs}[a]), typified by fairly low density ($\approx$ 1.5 \densunit{}) and a varying amount of electron cyclotron heating (ECH). A cryogenic Ar pellet is injected (Ar-PI) from the low-field side midplane to initiate the fast plasma shutdown and generate the REs \cite{Evans1998,Hollmann2010}. As shown in Fig. \ref{fig:ipfig} (magenta color), the conventional scenario injects the Ar pellet at \Ip{} = 1.2 MA (t = 1.2 s), and is typified by 1-2 keV average core temperature (\Teavg{}) as shown in the magenta data-points in Fig. \ref{fig:ipfig}).

In contrast, the high \Te{} scenario utilized for this work uses the same basic actuators with a different timing. The ECH power is applied slightly earlier (0.2 s vs 0.3 s) and the density feedforward target is reduced by 20\%.  This sustains a strongly reversed magnetic shear early in the discharge and allows an internal transport barrier (ITB) to form \cite{Strait1995,Wolf2003}. Varying ECH power (here from 0.5 - 2.3 MW) allows \Teavg{} in the core of the plasma to be scanned from 2-12 keV inside the ITB. A key difference is also in the timing of the Ar pellet injection, which arrives earlier (0.35 s or 0.7 s) to take advantage of the strong ITB phase. Waiting longer in these conditions results in a penetration of the ohmic current, a loss of reversed shear, and a loss of the ITB. As a result, \Ip{} is significantly lower in this high \Te{} scenario: 0.6 MA for the 0.35 s injection, and 0.8 MA for the 0.7 s injection. This will have important consequences when converting the RE beam (`plateau') current to the RE seed current, as will be described in Sec. \ref{sec:seed}. Note the highest \Teavg{} discharges produced REs without Ar PI, as will be described in Sec. \ref{sec:prompt}.

Equilibria corresponding to each \Ip{} have been reconstructed and are shown in Fig. \ref{fig:EQs}. Reconstructions are done with internal current profile and kinetic pressure constraints achieved via dedicated discharges with short NBI blips for motional Stark effect (MSE) and charge exchange recombination spectroscopy (CER). As can be seen the low \Ip{} case is found to have rather reversed magnetic shear, though even the conventional 1.2 MA scenario has not had time to fully relax to a stationary Ohmic current profile with q on-axis not yet unity.  The poloidal flux change from the edge (Fig. \ref{fig:EQs}[c]) is also presented, which as can be seen peaks on-axis from $\approx$ 0.8 to 1.8 V-s. This can be compared to the ITER value of 75 V-s \cite{Boozer2018}. Clearly, avalanche gain is a much smaller effect in these discharges as compared to fusion-grade tokamak plasmas, as will be discussed in Sec. \ref{sec:seed}

\begin{figure}
\centering
\includegraphics[width=0.4\textwidth]{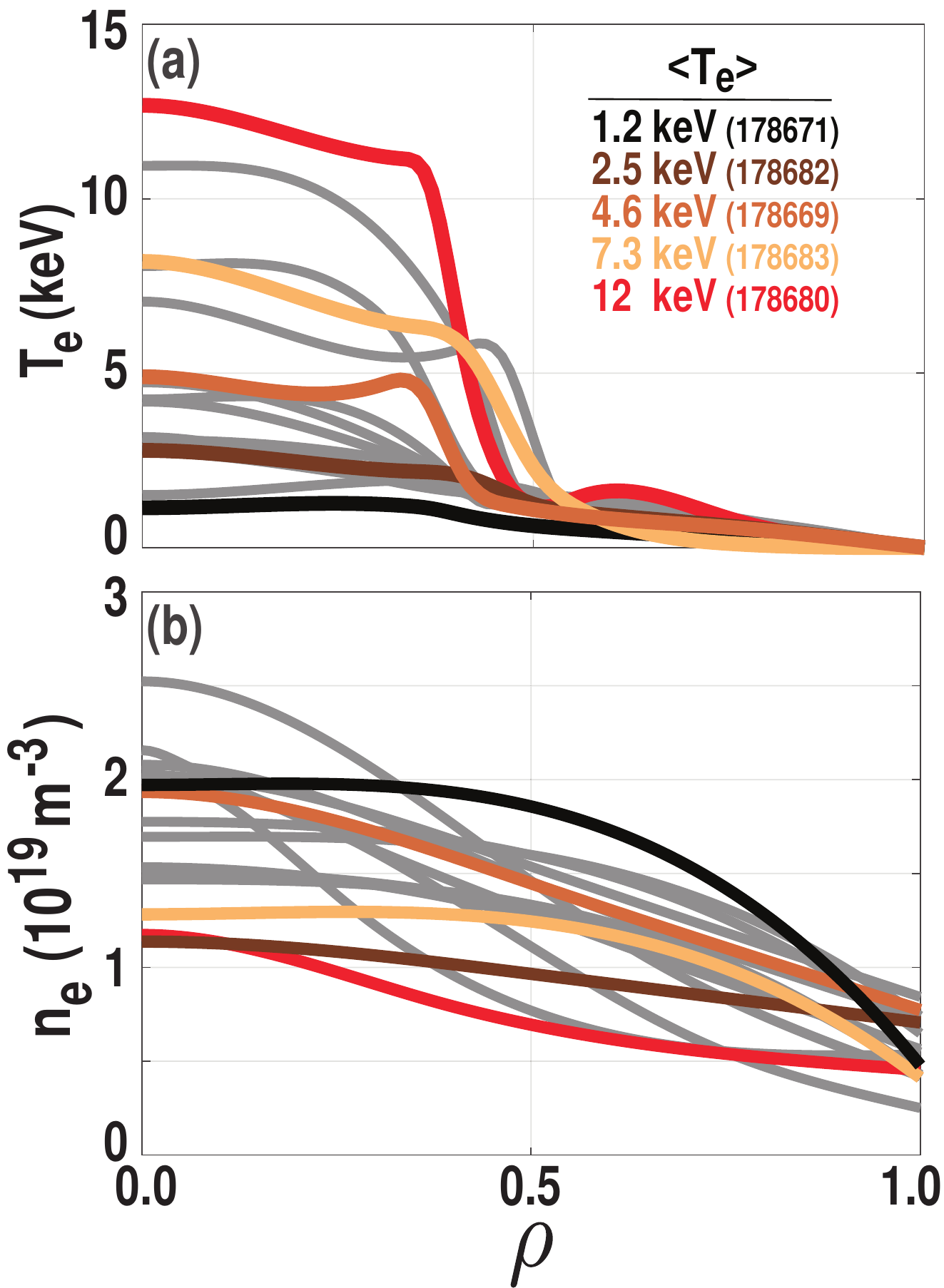}
\vspace{-10 pt}
\caption{Profiles of pre-disruption electron (a) temperature (\Te{}) and (b) density (\dens{}) accessed in this experiment. Discharges studied in detail later in this work are highlighted in color, while others are left in grey.}
\label{fig:profs}
\end{figure}

Pre-disruption electron temperature (\Te{}) and density (\dens{}) profiles for the discharges studied in this work are shown in Fig. \ref{fig:profs}. Profiles are fit using a combination of data from the Thomson Scattering, Electron Cyclotron Emission, and CO$_2$ interferometer diagnostics. As can be seen, the ITB exists only in the inner half of the radius, consistent with the location of minimum $q$ in Fig. \ref{fig:EQs}(b). A very sharp \Te{} gradient is observed at mid-radius consistent with ITB formation, and later in Sec. \ref{sec:pellet} the Ar ablation will be shown to be intense as the pellet enters the ITB. A few profiles are highlighted in color in Fig \ref{fig:profs}. These discharges, at \Teavg{} of 1.2, 2.5, 4.6, 7.3, and 12 keV will be highlighted in later detailed analysis and kinetic modeling. Note \Teavg{} is the average \Te{} over the central ITB region, $\rho$ = 0.0 to 0.35.


\section{Seed Current Extraction in a Seed Dominated Regime}
\label{sec:seed}

As this work focuses on the RE seed formation dynamics, the final RE beam `plateau' current (\Ire{}) has to be converted into the initial seed RE current prior to avalanche multiplication. Fortunately, this process is significantly simplified in these experiments owing to the very low values of \Ip{} used. Crudely, the avalanche gain factor is proportional to $\exp{(I_{P}/ (I_A\ln{\Lambda}))}$ \cite{Rosenbluth1997}, where taking $\ln{\Lambda}$ = 18, the Alfv\'en current $I_A$ = 17 kA, and \Ip{} = 0.6 MA discharge would be able to provide at most an avalanche gain factor of $\exp(2) = 7$. To simplify the extraction of the seed current for the remainder of the study only the \Ip{} = 0.6 MA discharges are discussed. This crude analysis will now be extended using more modern treatments to yield more quantitative estimates of the avalanche gain, though the main result will remain unchanged: owing to the low initial \Ip{} and large observed \Ire{}, the avalanche effect is small in these discharges and can be removed without introducing significant experimental uncertainty. For this reason, these plasmas can be thought of as existing in a RE seed dominated regime.

\begin{figure}
\centering
\includegraphics[width=0.4\textwidth]{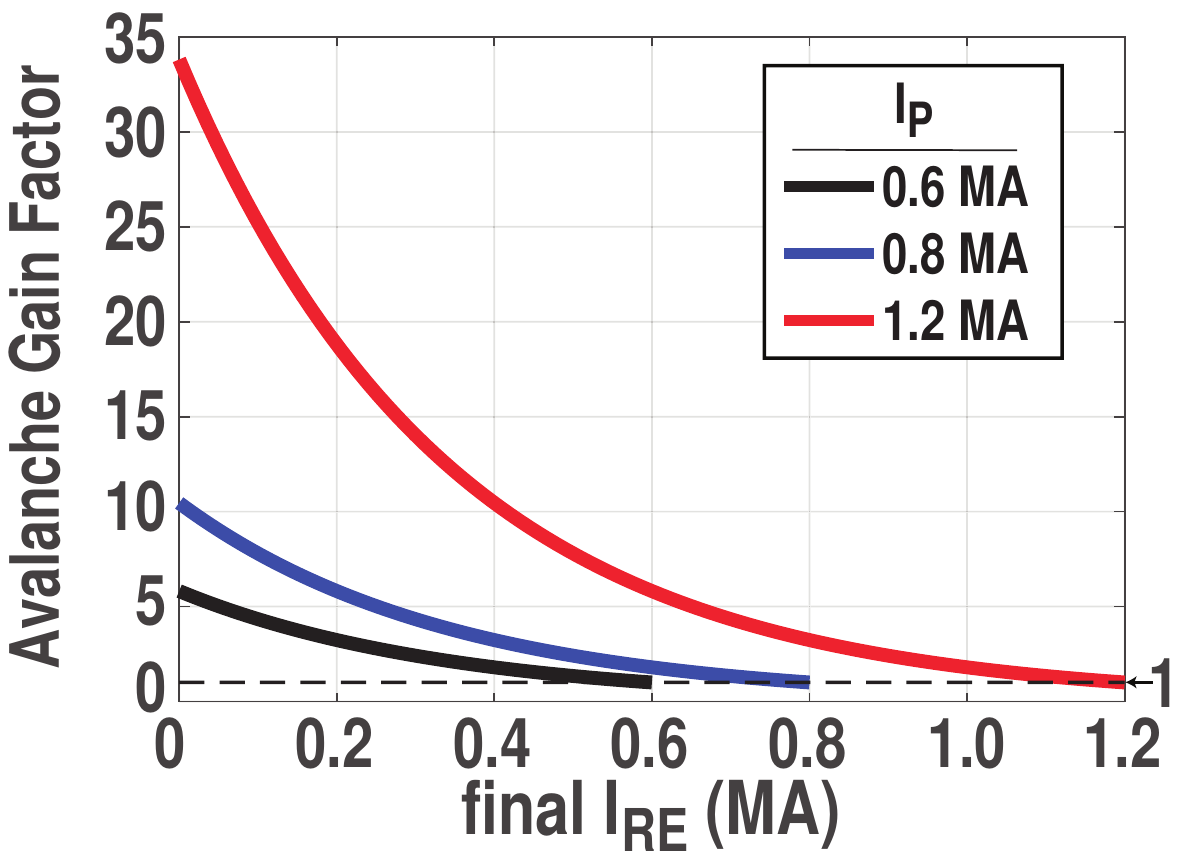}
\vspace{-10 pt}
\caption{The expected avalanche gain including the finite flux preseved in the RE current as a function of \Ip{} and \Ire{} \cite{Breizman2019}.}
\label{fig:aval}
\end{figure}

Figure \ref{fig:aval} indicates evaluations of the avalanche gain for the various \Ip{} scenarios using a 0-D analysis provided in Eq. 99 of Ref. \cite{Breizman2019}. A slight modification to the derivation is made, expanding the effective inductance from $\mu_0 R_0 \ell_i/2$ to $\mu_0 R_0(\ell_i/2 + \ln{(8 R_0/a)}-2)$. This is done to account for the magnetic energy available from outside the vacuum vessel which can contribute flux in DIII-D but not ITER due to the disparate vessel resistive times. This modification yields the following expectation for the avalanche gain, 
\begin{equation}
\text{Gain} = \exp{\left(\frac{2(\ln{(8 R_0/a)}-2 + \ell_i/2)}{\sqrt{Z+5} \ln{\Lambda}} \frac{I_\text{P} - I_{\text{RE}}}{I_\text{A}} \right)},
\label{eq:gain}
\end{equation}
where major radius $R_0$ = 1.67, minor radius $a$ = 0.6, internal inductance $\ell_i = 1$ (from measurement), charge state $Z = 5$ (estimated) and Coulomb logarithm $\ln \Lambda = 18$ (estimated), this gives the result shown in Fig. \ref{fig:aval}, varying both the initial \Ip{} and the final \Ire{}. When initial \Ip{} = final \Ire{} (full conversion), there can be no avalanche gain as no flux has been consumed (Gain Factor = 1).

\begin{figure}
\centering
\includegraphics[width=0.4\textwidth]{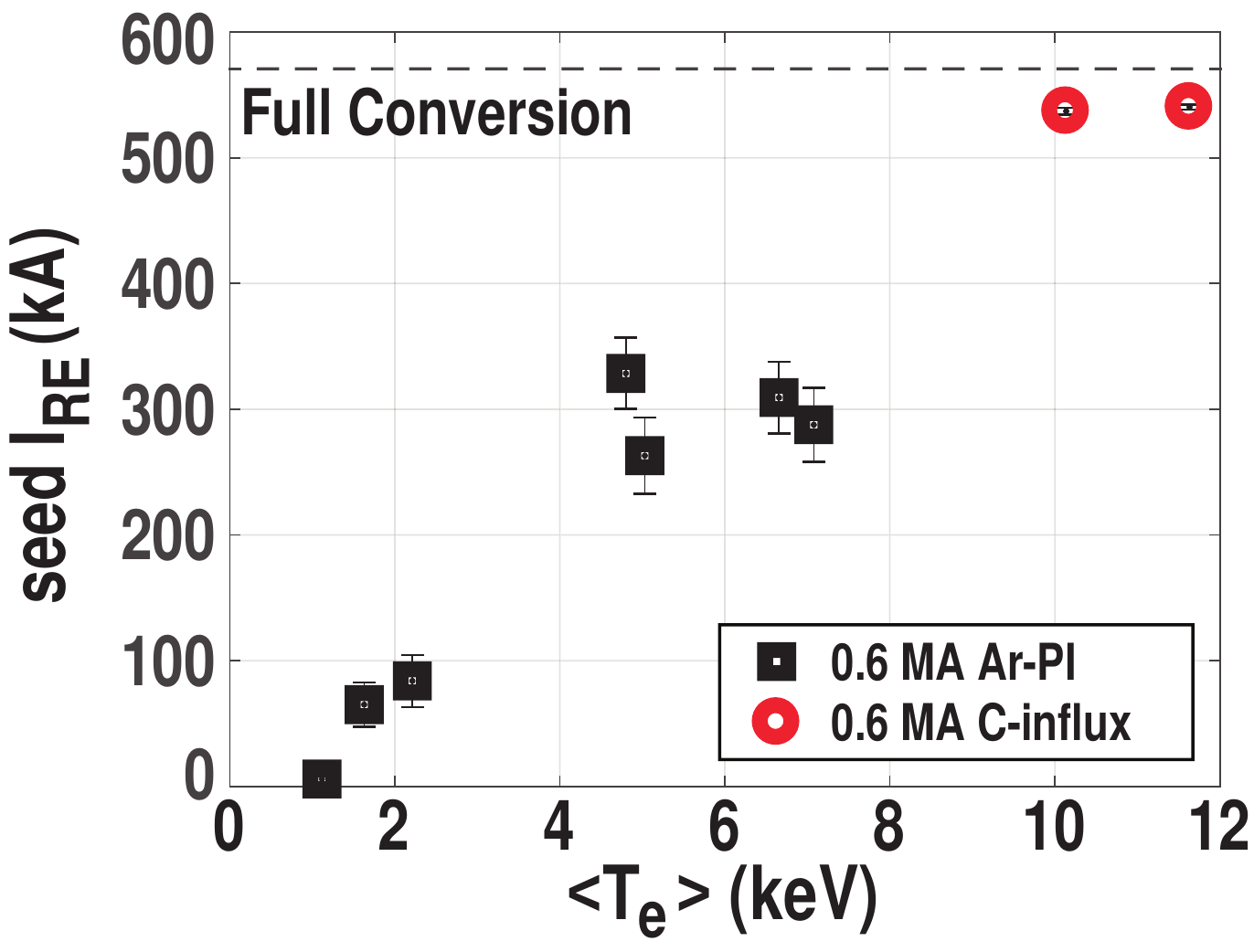}
\vspace{-10 pt}
\caption{Inferred seed RE current (dividing final \Ire{} by estimated avalanche gain) as a function of pre-disruption \Teavg{} for \Ip{} = 0.6 MA plasmas.}
\label{fig:Iseed}
\end{figure}

In these experiments, the initial \Ip{} and the final \Ire{} are measured experimentally, allowing extraction of an estimate of the avalanche gain factor via Eq. \ref{eq:gain}. This factor, ranging from 1 to 5, is applied to the observed final \Ire{} to obtain Fig. \ref{fig:Iseed}, an experimental assessment of the seed \Ire{} variation with \Teavg{}. As the gain factor is rather small, the alternate path of estimating the seed \Ire{} from pellet ablation light as in Ref. \cite{Hollmann2017} is not here pursued. A dedicated future study is planned, as the analysis procedure is very challenging at high \Teavg{} due to the dramatically different pellet dynamics as will be described in Sec. \ref{sec:pellet}. It should also be noted that the enhanced avalanche effect due to partial screening \cite{Martin-Solis2015,Hesslow2019} is not included in these estimates. However, as the effect becomes stronger with the electric field, it may not be a large correction in these low \Ip{} plasmas.


\section{Pellet Ablation and Thermal Quench Dynamics}
\label{sec:dynamics}

In this section the dynamics resulting from the variation in \Teavg{} on the observed Ar pellet ablation, thermal quench duration, and MHD behavior will be described. Significant variations with \Teavg{} are observed in Ar pellet ablation and TQ duration, demonstrating that consideration of the \Te{} effect in isolation of these other dynamics misses key interactions. No clear variation of the MHD dynamic with \Teavg{} is found.

\subsection{Pellet Dynamics}
\label{sec:pellet}

Fast shutdowns in all discharges with \Teavg{} $\leq$ 10 keV are intentionally initiated with the use of an injected Ar cryogenic pellet. The cylindrical pellet travels at 200 m/s and has dimensions 3 mm diameter x 3 mm length, and contains 4.9 $\times 10^{20}$ Ar atoms (1.9 Pa m$^3$ or 14 Torr-L). The 3 mm pellet crosses a fixed point in space in 15 $\mu s$, which will be shown to be shorter than the TQ duration. Measurement of pellet ablation light (Ar-I @ 696.5 nm) is provided by an absolutely calibrated fast framing camera (18 kfps) observing the pellet trajectory \cite{Yu2013,Hollmann2017}.

\begin{figure}
\centering
\includegraphics[width=0.4\textwidth]{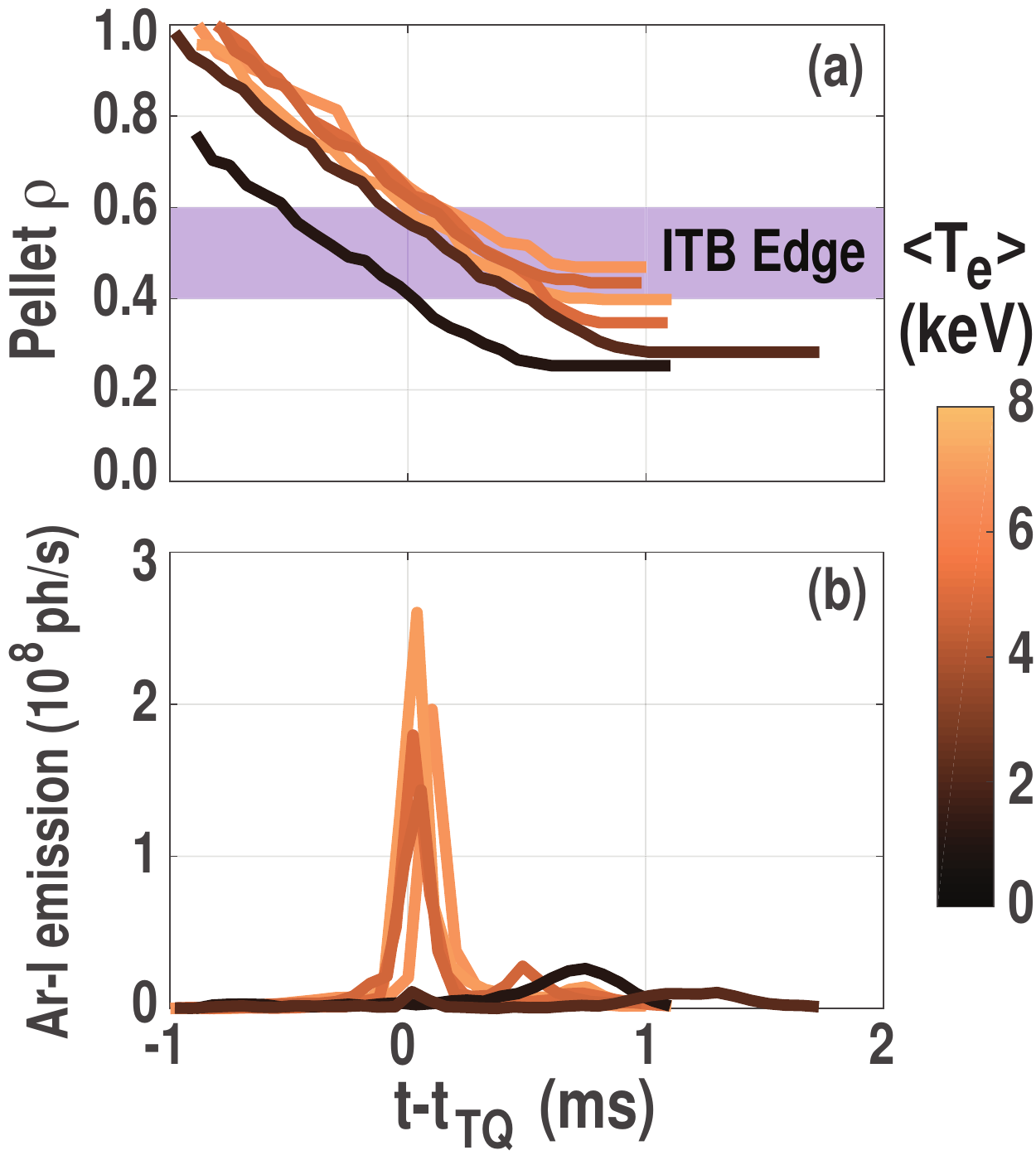}
\vspace{-10 pt}
\caption{Estimation of (a) pellet trajectory and (b) measured total Ar-I ablation light for a subset of discharges, colored by \Teavg{}. As \Teavg{} rises, the pellet ablation becomes intense and prompt, roughly when the pellet enters the ITB edge.}
\label{fig:pellet1}
\end{figure}

Estimation of the pellet trajectory from initial pellet position and velocity alongside total measured Ar-I emission is shown in Fig. \ref{fig:pellet1}. Even from the total ablation light dramatic differences from low to high \Teavg{} are observed. At low \Te{}, the pellet is only partially ablated by the intial thermal plasma, and survives to produce further ablation from the RE seeds themselves well after the TQ. This scenario is equivalent to the detailed study of Ref. \cite{Hollmann2017}, where the post-TQ ablation phase is used to estimate the RE seed.

At high \Te{}, the pellet ablation light is far more intense and prompt, with nearly all ablation coincident with the TQ. The intense ablation light appears at a consistent time with the pellet entering the ITB edge. Clearly, the high \Te{} of the ITB is extremely effective at ablating the pellet, as might be expected by thermal ablation which scales like $T_e^{5/2}$. 



\begin{figure}
\centering
\includegraphics[width=0.4\textwidth]{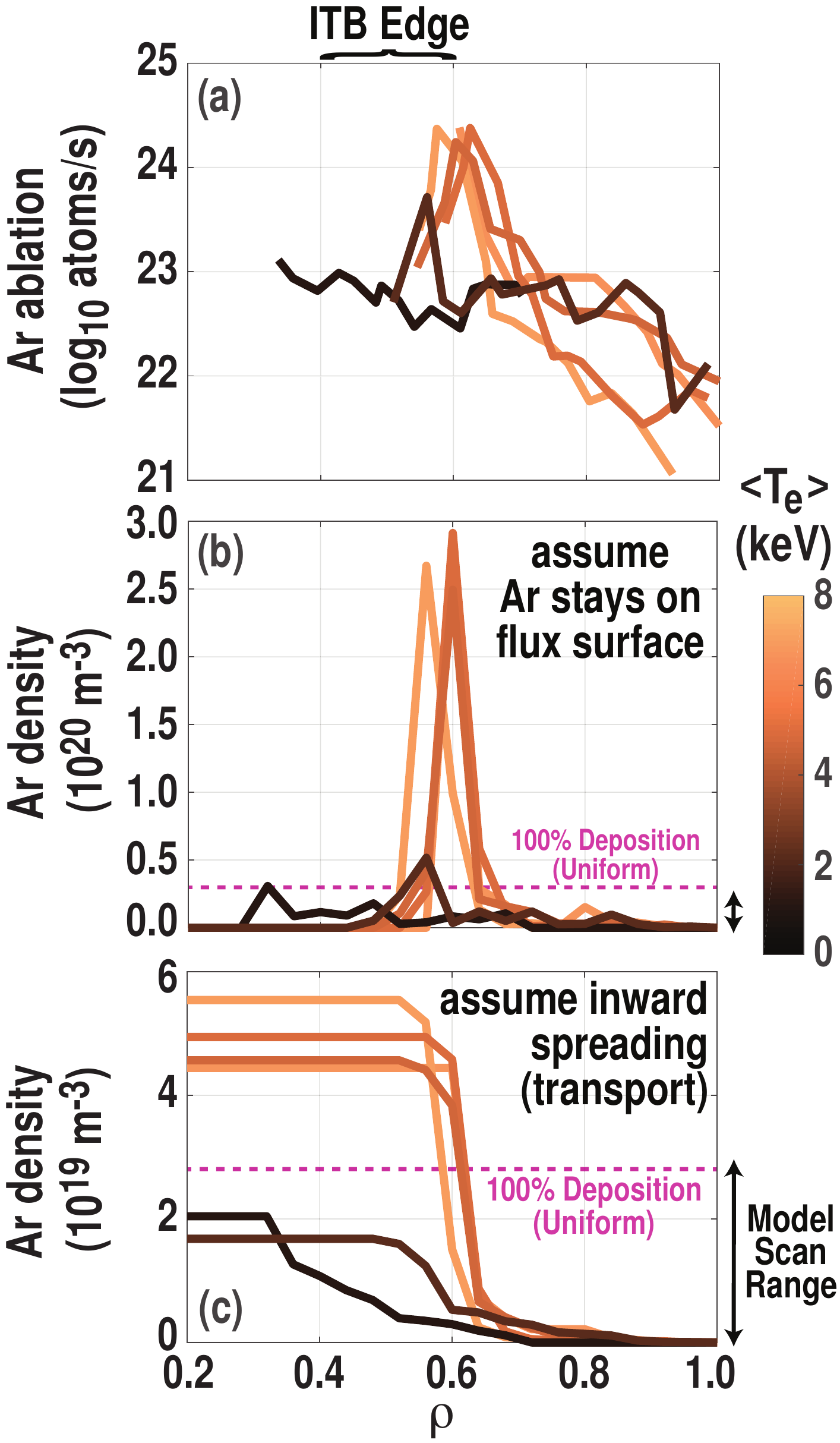}
\vspace{-10 pt}
\caption{Inference of deposited Ar profile from Ar-I ablation light. (a) Ar ablation rate vs radius, (b) Ar density assuming zero radial transport (local deposition), (c) Ar density assuming an inward spreading, effectively spreading the Ar throughout the ITB region. Also included is the uniform deposition limit (magenta) and the scan range of the kinetic modeling described in Sec. \ref{sec:model}.} 
\label{fig:pellet2}
\end{figure}

Measurement of the calibrated Ar-I ablation light allows determination of the ablated Ar quantity and with some assumptions its distribution throughout the profile. The Ar radial profile will be used for kinetic modeling of the RE seed formation in Sec. \ref{sec:model}. Figure \ref{fig:pellet2}(a) presents the rate of ablated Ar atoms, scaled such that the total ablation is equal to the pellet inventory. Data only up to 0.25 ms after the TQ is presented, to remove contributions from any post-TQ RE seed induced ablation. Post-TQ dynamics are beyond the scope of this manuscript, though they are briefly discussed in \ref{sec:gri}. Consistent with Fig. \ref{fig:pellet1}(b), the Ar quantity is over one order of magnitude higher in the high \Teavg{} cases.  Application of a published pellet thermal ablation model \cite{Parks1998,Parks2012} using the profiles of Fig. \ref{fig:profs} under-predicts this ablation rate significantly.  Application of a forthcoming pellet thermal ablation model \cite{Parks2020} improves on the agreement, finding the pellet can only penetrate 0.1 in $\rho$ units but only if the initial \Teavg{} of 8 keV is used (ignoring the cooling front and TQ). Any deficit from thermal ablation models indicates either significant ablation from RE seeds or more simply inaccuracies in the thermal ablation model. Future work is planned to quantitatively compare existing and new ablation models using this experimental data, and to estimate the RE seed using the techniques of Ref. \cite{Hollmann2017}.


Several ways of estimating the Ar density profile from the Ar ablation are shown in Fig. \ref{fig:pellet2}(b,c). Firstly, the simplest estimate is deposition of the full pellet inventory uniformly across the plasma. This gives an Ar density (\nAr{}) of 0.29\nArunit{}, as indicated by the dashed magenta lines in Fig. \ref{fig:pellet2}. The second estimate assumes zero radial transport of the Ar impurity. Here the Ar density is simply the integrated Ar ablation divided by the area of the plasma annulus in which it is deposited. At high \Te{} this results in extremely large localized Ar density at the ITB edge and zero inside. The third estimate assumes a full inward radial spreading. As the ablation is at the ITB edge, this assumption functionally homogenizes the inside of the ITB for the high \Teavg{} cases. The motivation for this is to mimic stochasticity within the ITB region, where the Ar would spread inwards and the thermal plasma transport outwards. Note in both assumptions the low \Te{} \nAr{} profile is below the 100\% uniform deposition limit - this is because the pellet is not completely ablated at low \Teavg{} until significantly after the TQ. The \nAr{} scan range used in later kinetic modeling is also indicated.


\subsection{Thermal Quench Dynamics}
\label{sec:TQ}

\begin{figure}
\centering
\includegraphics[width=0.48\textwidth]{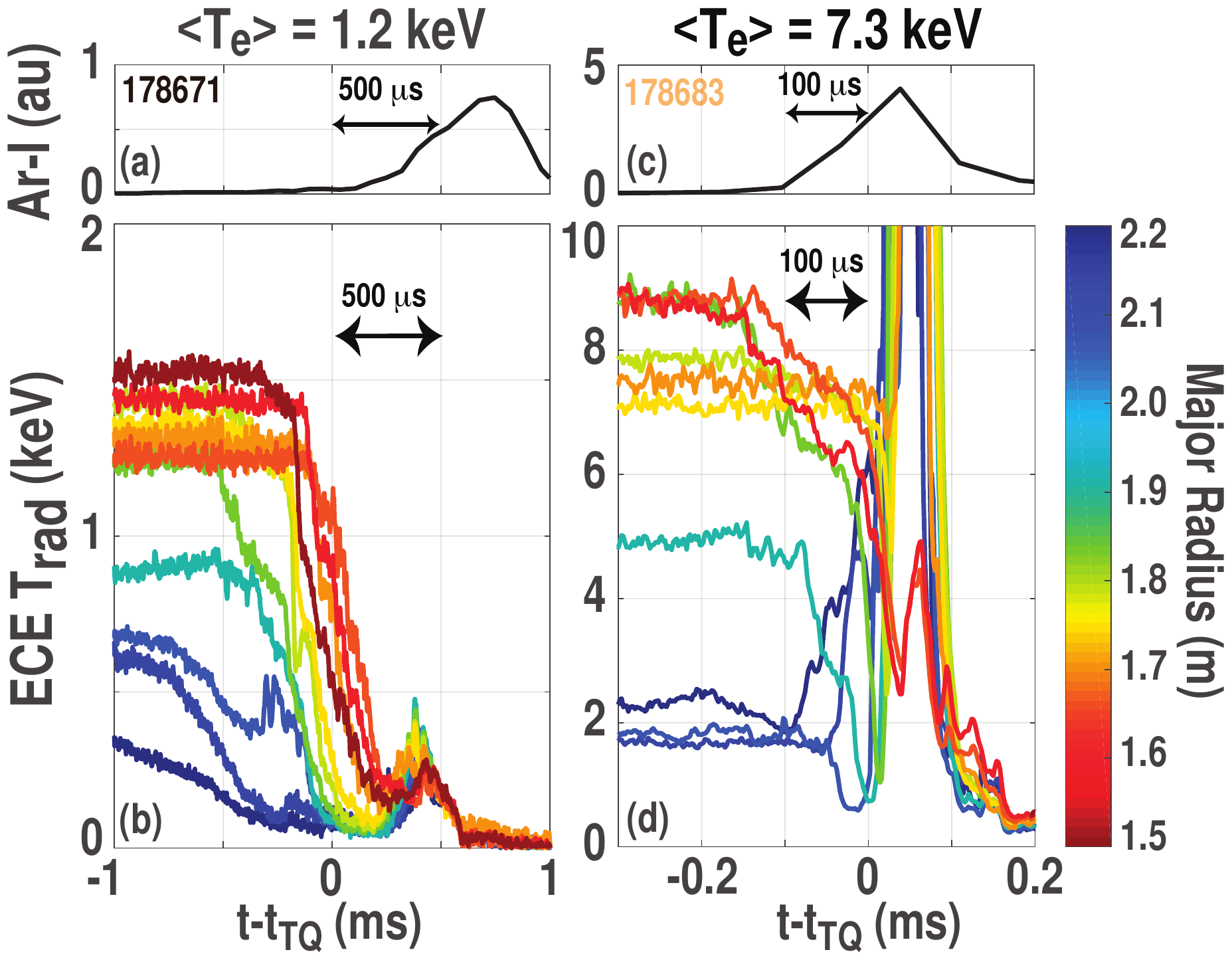}
\vspace{-10 pt}
\caption{Fast dynamics of the TQ in low (left) and high (right) \Te{} plasmas. Ar-I emission via fast camera measurement and \Te{} from ECE are displayed vs. time. Note the time axis is zoomed for the high \Te{} example.}
\label{fig:TQzoom}
\end{figure}

The impact of the intense pellet ablation on the thermal quench (TQ) dynamics is now shown to occur simultaneously with a shorter TQ duration and enhanced RE seed production. Candidate mechanisms to explain the shorter TQ are then discussed. The main measurement is the ECE radiometer, sampled every 2 $\mu$s. Care is taken to only consider discharges with no evidence of density cutoff.

Discharges at high and low \Teavg{} are compared in Fig. \ref{fig:TQzoom}, along with the Ar-I ablation light. As can be seen the entire TQ dynamic is significantly accelerated at high \Teavg{}. This is striking considering that it should take significantly longer to cool a high \Te{} plasma at constant impurity content. At low \Te{}, a large fraction of the Ar ablation light is delayed from the main TQ, indicating that only a minority of the pellet inventory played a role in the TQ dynamics. Further Ar-I light observed after the TQ is due to additional Ar pellet ablation by the growing RE population. The post-TQ REs are observed to also drive some ECE after the TQ. At high \Te{}, the entire pellet is ablated over a very short duration, and the ECE emission almost immediately shows signs of significant non-thermal activity (broadband spikes), which further could in principle drive further Ar ablation. As discussed in Sec. \ref{sec:pellet}, published thermal ablation models underpredict the pellet ablation \cite{Parks1998,Parks2012} though using an upcoming model (ignoring the cooling front and TQ) improves the agreement \cite{Parks2020}.

The observed ablation enhancement and shortened TQ observed at high \Te{} is unexpected. Three mechanisms are now proposed to explain these results. The first explanation is that the observed ablation is purely due to conventional thermal bulk ablation, and that any mismatch with experiment is simply due to gaps in these models.  Detailed follow-on work is planned to benchmark existing and upcoming ablation models using this high \Te{} data.

A second mechanism involves the RE seeds themselves playing an important role in the ablation. A conceptual picture of this event is that a vicious cycle is set up, whereby increased Ar ablation begets faster cooling (via radiation) which begets more RE production which begets more Ar ablation, and so on. This mechanism requires radial transport to repopulate the flux surface with energy and bulk electrons, and also poloidal transport to quickly distribute the Ar impurity throughout the flux surface. This picture would be assisted by strong stochasticity within the core plasma, enabling the aforementioned transport. The net outcome of this mechanism would be a prompt ablation of the pellet alongside strong RE seed production, which is consistent with experiment.

A third mechanism is that the enhanced rate of cooling is due to parallel transport across stochastic fields as the ITB collapses. The rate of parallel transport scales like $T_e^{7/2}$ \cite{Braginskii1965}, thus all else being equal parallel transport should accelerate the TQ as \Te{} rises. Note that the above mechanisms may co-exist and act together to accelerate the TQ.

\begin{figure}
\centering
\includegraphics[width=0.4\textwidth]{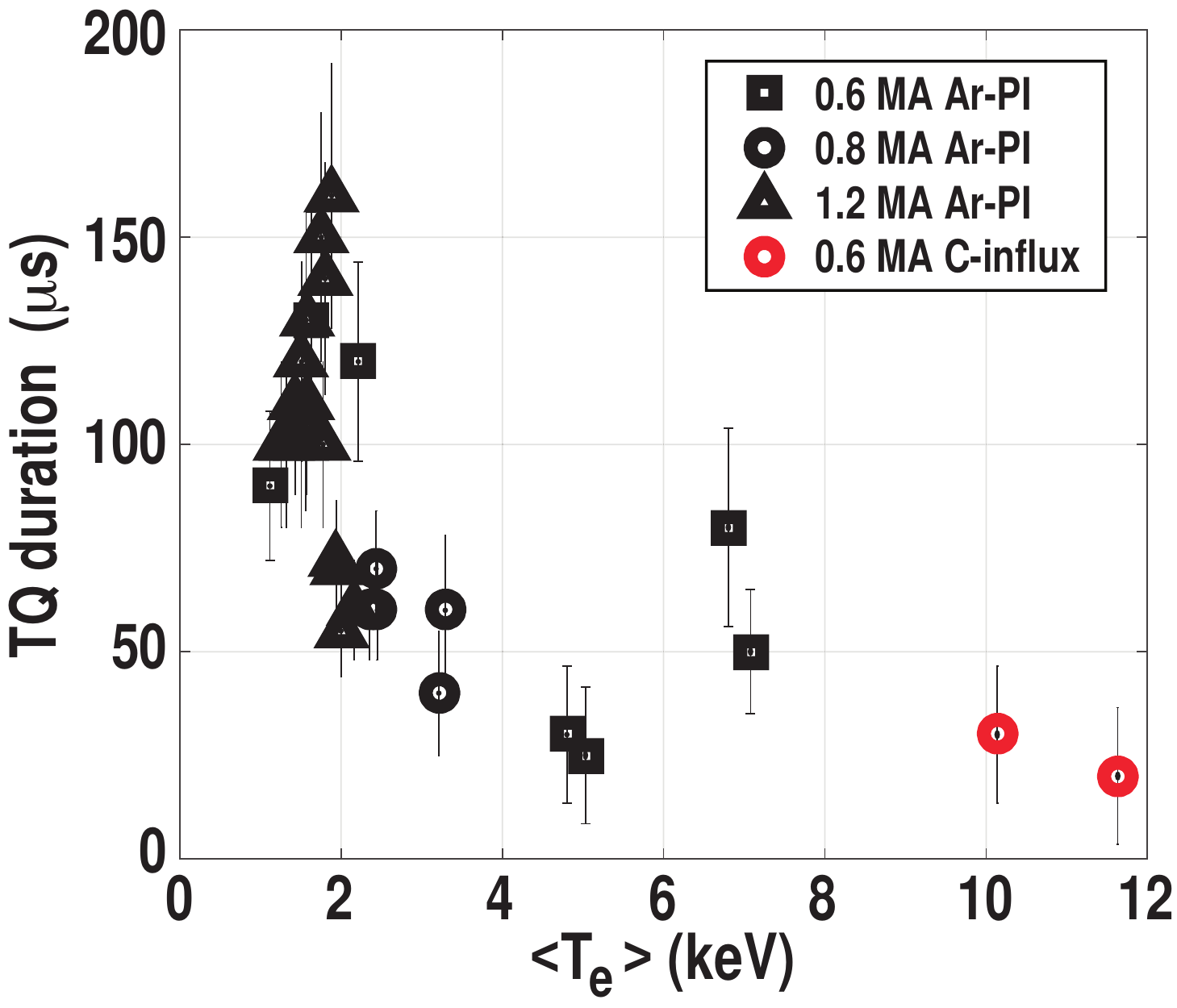}
\caption{Estimate of the TQ duration for the discharges of this study as a function of \Teavg{}, obtained from ECE. A single aggregate time is inferred from multiple ECE channels approximating the time to cool from 80 to 20\% of the initial \Te{}, ignoring any slow (pre-TQ) cooling. Due to the complexity of the TQ dynamic, this aggregate time should be treated as approximate.}
\label{fig:TQdur}
\end{figure}

Figure \ref{fig:TQdur} shows a rough estimate of the experimental dependence of the TQ duration on \Teavg{}. Note extraction of the TQ duration is challenging, with variation possible across the plasma radius. These values are extracted by inspection of the ECE measurement, as in Fig. \ref{fig:TQzoom}, thus some uncertainty is expected. A single aggregate time is inferred from multiple ECE channels approximating the time to cool from 80 to 20\% of the initial \Te{}, ignoring any slow (pre-TQ) cooling. Despite measurement challenges, the trend of decreasing TQ duration with \Teavg{} is clear and robust. Furthermore, kinetic modeling will be presented in Sec. \ref{sec:model} with the TQ duration scanned as an independent variable.

It should be noted that while these TQ durations are faster than commonly used in ITER modeling (0.5 - 1.0 ms \cite{Martin-Solis2017}), the dominant scaling of the TQ duration is expected to be the minor radius. With DIII-D being $\approx$ 4x smaller in linear dimension than ITER, experimental values on DIII-D are in line with the aforementioned ITER modeling. Furthermore, observed TQ durations on JET in high \Teavg{} ITB conditions were found to be below 0.1 ms, also roughly consistent with these DIII-D observations when scaling by minor radius \cite{Riccardo2005}.

Fig. \ref{fig:TQdur} (together with Figs. \ref{fig:ipfig} and \ref{fig:pellet1}[b]) should be seen as a key empirical result of this study. While the experimental actuation was on \Teavg{} via ECH heating, the impact of the higher \Teavg{} indirectly accelerated the TQ duration via either enhanced Ar ablation/radiation or parallel transport. The combination of high pre-disruption \Te{} and faster TQ conspires to produce prodigious RE production in experiment, and illustrates the limitation of taking TQ duration as a constant input parameter.

\subsection{MHD Instability Dynamics}
\label{sec:MHD}

\begin{figure}
\centering
\includegraphics[width=0.4\textwidth]{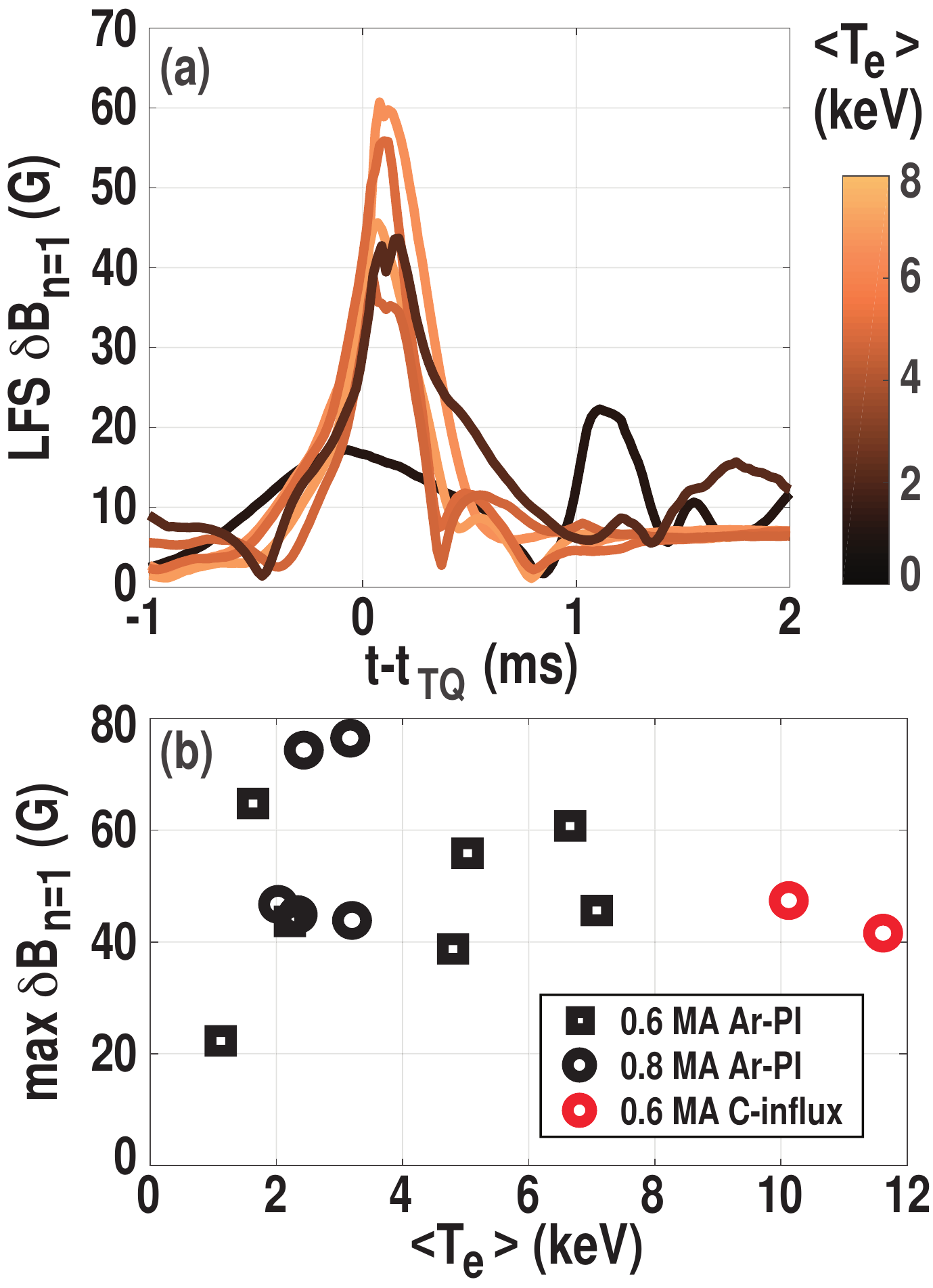}
\vspace{-10 pt}
\caption{(a) Dynamics of the non-axisymmetric magnetic field ($\delta B$) during the TQ plotted vs. time and (b) max $\delta B$ vs. \Teavg{}. No clear \Teavg{} dependence is found with the external magnetics.}
\label{fig:MHD}
\end{figure}

Differences in the MHD dynamics have been previously invoked to explain differences in RE seed formation as magnetic configuration is varied, and going from large to small devices \cite{Izzo2011,Izzo2012}. The dependence of the measured MHD dynamics on \Teavg{} is shown in Fig. \ref{fig:MHD}. The n=1 $\delta B$ poloidal component is extracted from a toroidal array of Mirnov probes on the low-field side midplane on DIII-D \cite{King2014}. No systematic dependence of the observed n=1 $\delta B$ is found, and peak $\delta B$ is around 40-60 G for all \Teavg{}.  Note that at the measurement location, the toroidal field is around 1.6 T, so $\delta B / B$ is around 3 $\times 10^{-3}$.  

While MHD-induced RE losses effects are likely important overall, they appear to not be changing with \Teavg{}. One caveat is that some instabilities (such as an internal mode) may not couple strongly to external magnetics, and thus cannot be resolved. Finally, an \Ip{} spike was always observed in these discharges with the exception of the coldest case (that also had the smallest $\delta B$). No trend in \Ip{} spike size with \Teavg{} was observed. 


\section{Kinetic Model Interpretation}
\label{sec:model}

Experimental measurements are interpreted using a recently developed kinetic models of the RE seed formation process \cite{Aleynikov2017}. This model self-consistently treats the cooling of a bulk Maxwellian electron distribution via impurity radiation, the generation of the induced electric field, and the formation of the RE seed population. This model improves over Ref. \cite{Smith2008a} by solving a time-dependent Fokker-Planck kinetic equation modelling the evolution of the hot electron population self-consistently with the power-balance equation and the equation for the electric field. The parallel current density (\Jpar{}) is kept constant, and as such the kinetic model treats the TQ but not the CQ. The model is local (0-D in space), and as such must be independently used at each radial position in an experimental profile. The main limitation of this model is the absence of radial transport of any kind, owing to its 0-D formulation. The impurity density is assumed to be uniform over the flux surface, and the effects of the pellet cloud expansion, discussed in Ref. \cite{Aleynikov2019}, are therefore ignored.

Inputs to this model are the local values of \Te{}, \dens{} (Fig. \ref{fig:profs}) and \Jpar{} (Fig. \ref{fig:EQs}) across the profile for a given discharge, and the Ar density profile (\nAr{}, which sets the TQ duration in the model). In its application to understanding these experiments, the model is independently evaluated across the radius for the experimental profiles of the colored shots in Fig. \ref{fig:profs}, and \nAr{} is used as the independent variable. Outputs of the model are the RE seed density (\nre{}), the RE seed current density (here expressed as a \% of the pre-disruption current density), the TQ duration, and the mean seed RE energy (\Ere{}). Aggregate metrics such as the total RE seed current can then be obtained by integrating the radial profile. As a side-note, the kinetic model also allows estimation of the Dreicer production \cite{Dreicer1959,Dreicer1960} during the TQ, which is found to be small as the ratio of the induced electric field to the Dreicer field stays below around 1 \% for all conditions, as such Dreicer production is ignored in this work. Dreicer was also independently found to be negligible in the low \Teavg{} cases studied in Ref. \cite{Hollmann2017}.

\begin{figure*}
\centering
\includegraphics[width=0.68\textwidth]{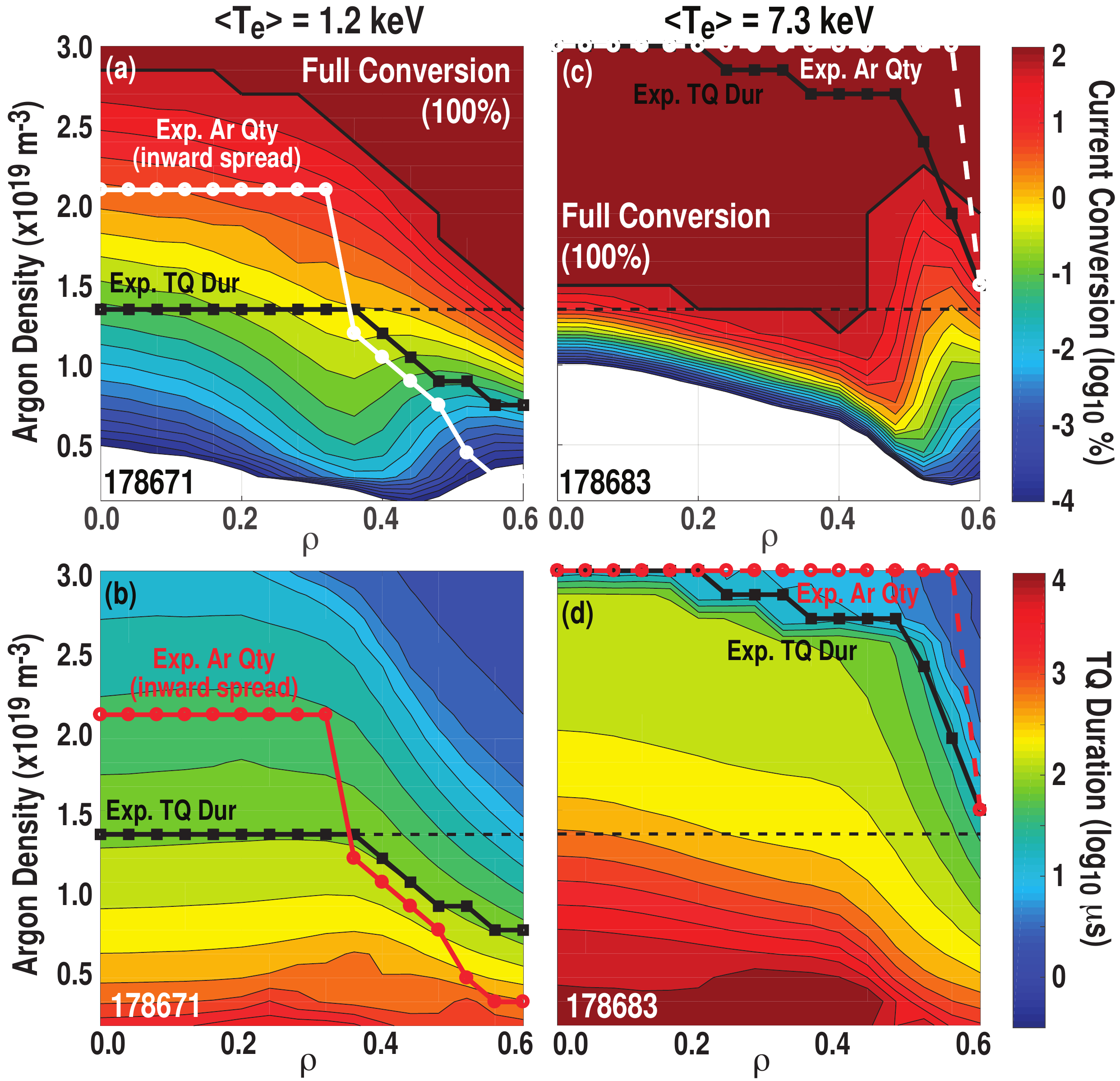}
\vspace{-10 pt}
\caption{Kinetic model predictions of the RE current generation (a,c) and TQ duration (b,d) across the radius and as Ar density is scanned for experimental discharges at low \Te{} (a,b) and high \Te{} (c,d). Dashed lines over-plotted are estimates of \nAr{} from Ar-I light (with inward spreading) and from matching the TQ duration.}
\label{fig:contours}
\end{figure*}

Example output of the kinetic model for low and high \Teavg{} discharges is shown in Fig. \ref{fig:contours}. The output of the current conversion (Fig. \ref{fig:contours}[a,c]) and the TQ duration (Fig. \ref{fig:contours}[b,d]) are displayed across the radial profile and as the Ar density (\nAr{}) is scanned. For the remainder of this section, different estimates for \nAr{} will be used to estimate the RE seed current and compare to experiment. Several of these \nAr{} estimates are indicated in Fig. \ref{fig:contours}: a constant \nAr{}, \nAr{} from experimental Ar-I light (assuming inward spreading), and \nAr{} from the experimental TQ duration. Since the model self-consistently treats the radiation cooling from the Ar impurity, the TQ duration is directly related to \nAr{} at a given radial point. Since the TQ duration is only characterized by a single number in this study, it can be locally matched across the profile by selecting the appropriate \nAr{} to yield the desired TQ duration. Simply put, matching the experimental TQ duration is an alternate means of prescribing an Ar density in the kinetic model.

\begin{figure*}
\centering
\includegraphics[width=0.98\textwidth]{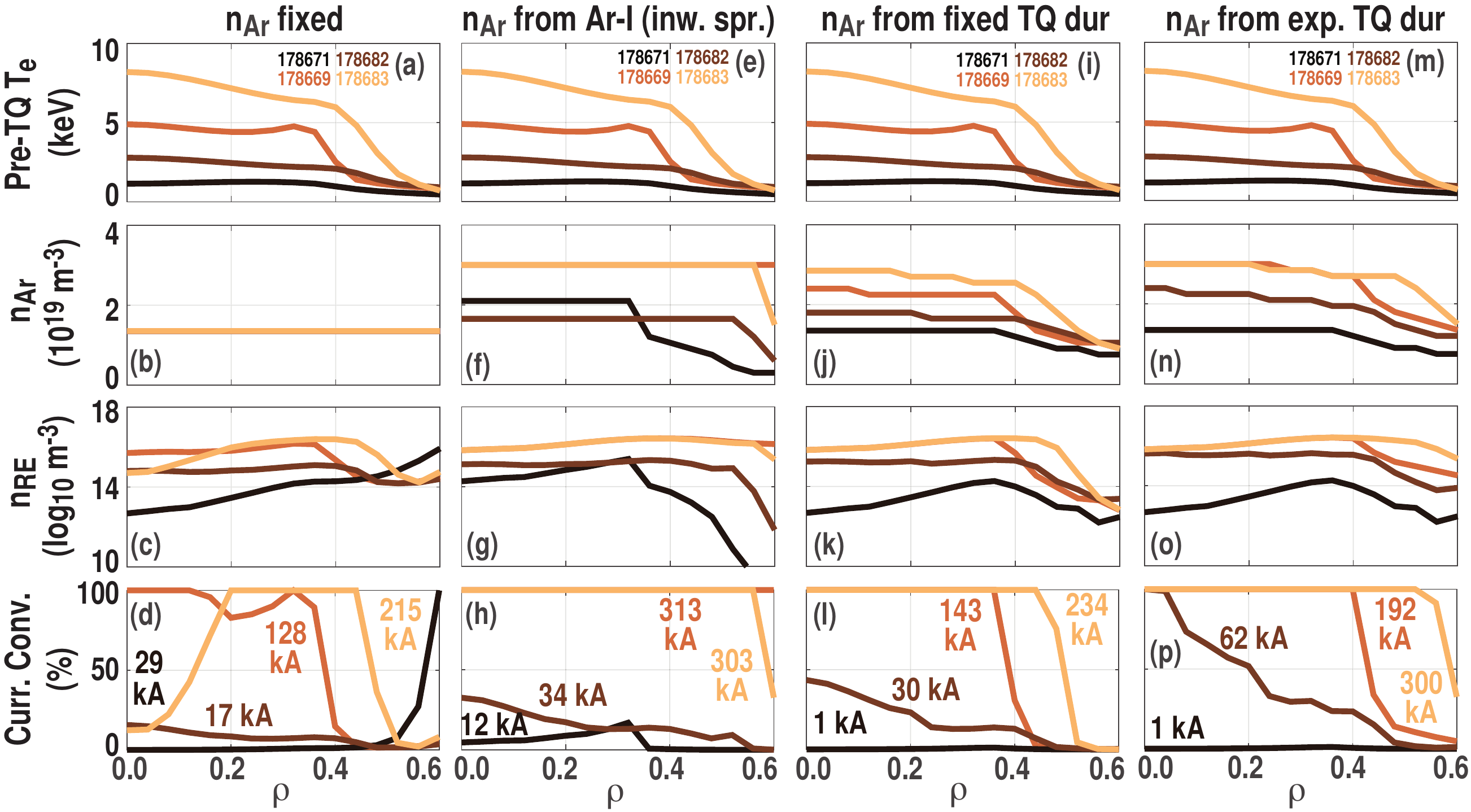}
\vspace{-10 pt}
\caption{Kinetic model predictions of seed RE production with several estimates of the Ar density. Panels (a-d) are with a fixed \nAr{} of 1.4 \nArunit{}, (e-h) are with \nAr{} estimated from Ar-I emission in Fig. \ref{fig:pellet2}(c), (i-l) are with \nAr{} fit to match at TQ duration of 125 $\mu$s, and (m-p) are with \nAr{} fit to match the experimental TQ duration in Fig. \ref{fig:TQdur}. Panels (a,e,i,m) are of pre-disruption \Te{} input to the model, (b,f,j,n) are of Ar density input to the model (\nAr{}), (c,g,k,o) are of output seed RE density (\nre{}), and (d,h,l,p) are of output current conversion from thermal to seed RE current. Integrated seed \Ire{} is also indicted in panels (d,h,l,p).}
\label{fig:modelAr}
\end{figure*}

Figure \ref{fig:modelAr} presents model predictions of the seed RE production across the plasma radius with several estimates of the Ar density. These are various 1-D cuts of the same data shown in Fig. \ref{fig:contours}.

First, the isolated effect of increasing \Teavg{} at fixed \nAr{} is considered, as shown in Fig. \ref{fig:modelAr}(a-d). At low \Teavg{}, the RE seeds are predicted to be strongest at the edge, although by examining Fig. \ref{fig:contours}(c) it can be seen that the radial position of most efficient RE production depends sensitively on \nAr{}. As \Teavg{} rises, the seed RE production shifts to the core and the seed RE current actually drops (Fig. \ref{fig:modelAr}[d]), indicating a non-monotonic behavior with \Te{} as described in Ref. \cite{Aleynikov2017}. This effect however is only found when \nAr{} is kept constant and is not observed in the experiment. As \Te{} further increases, the seed RE production moves towards occupying more of the plasma radius, and the seed \Ire{} increases. At the highest \Te{}, RE production is predicted to be in an annular region at mid-radius, again further evidence of non-monotonicity in \Te{} which relates to non-monotonicity in the radiative cooling rate of Ar as a function of \Te{}.

Next, similar data is shown in Fig. \ref{fig:modelAr}(e-h) for the experimental \Te{} profiles (Fig. \ref{fig:profs}) and experimental \nAr{} (Fig. \ref{fig:pellet2}(c)). Note that if the extracted \nAr{} is above the scan range, a value of 3 \nArunit{} is used. As shown in Fig. \ref{fig:contours}, this is enough Ar available to convert the entire current to RE current. The predicted seed \Ire{} in this case monotonically rises with \Teavg{}, as in experiment. The predicted seed RE production is now also strongest in the core, as opposed to at mid-radius. Unfortunately no existing diagnostic is available to constrain the radial profile of the seed RE production and is a key focus of diagnostic development at DIII-D.

As described, a different method of using the kinetic model to predict the seed \Ire{} is to infer \nAr{} from the TQ duration. These two quantities are directly linked in the model since the cooling is provided entirely by impurity radiation. A TQ duration of 125 $\mu$s is enforced across \Teavg{} and across the radius, giving the predictions shown in Fig. \ref{fig:modelAr}(i-l). With fixed TQ duration, the RE seed production now increases with \Teavg{} and is again found first in the core of the plasma. The experimental TQ duration is used in Fig. \ref{fig:modelAr}(m-p), and since faster TQ duration begets more effective RE production, now the RE seed rises more rapidly with \Teavg{}, and again begins in the core.

\begin{figure}
\centering
\includegraphics[width=0.48\textwidth]{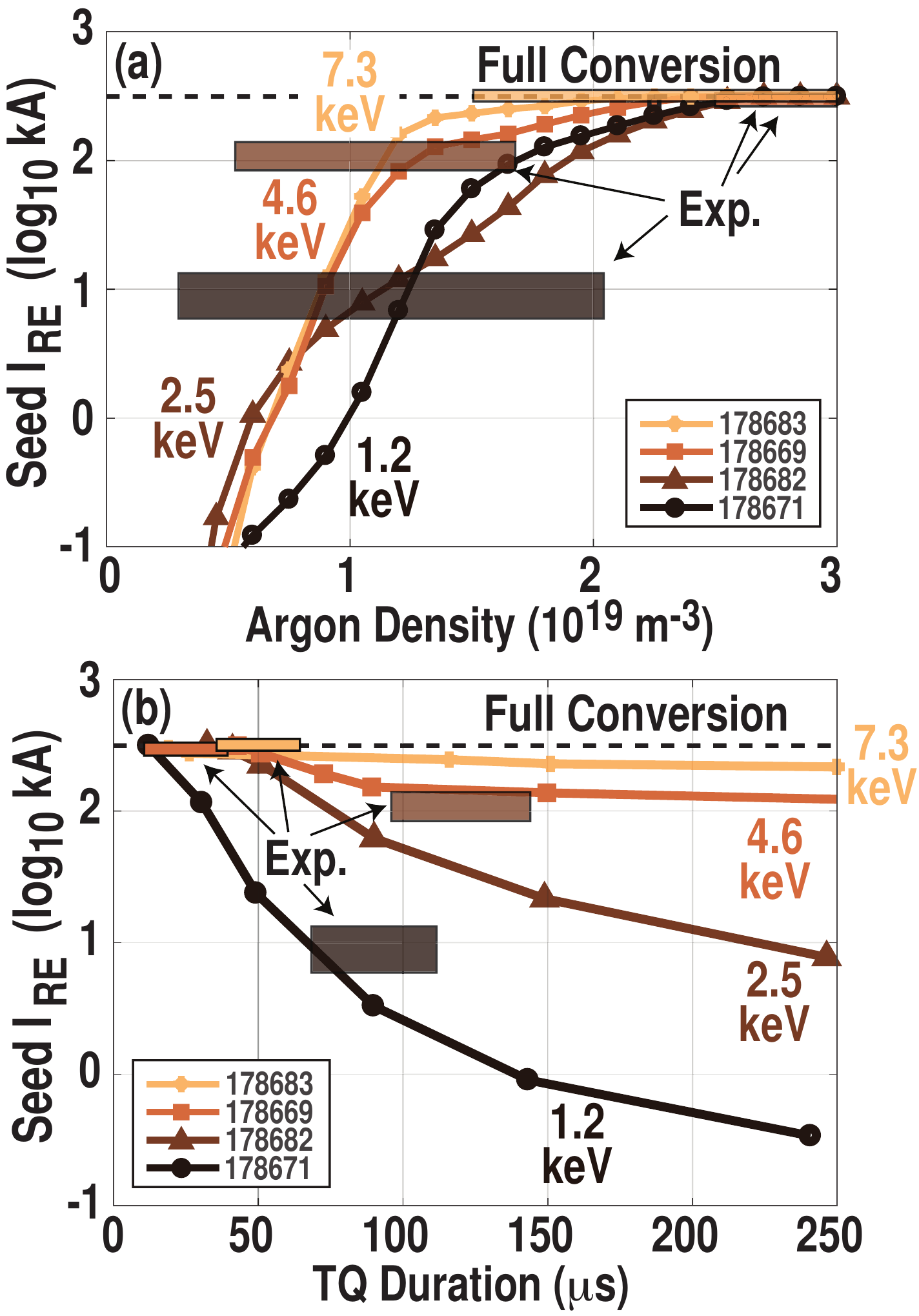}
\caption{Parametric dependence of the seed RE current on (a) uniform Ar density and (b) TQ duration, alongside experimental values given as ranges with a rough confidence interval.}
\label{fig:model1}
\end{figure}

1-D profile results presented in Fig. \ref{fig:modelAr} are summarized in Fig. \ref{fig:model1} by extracting the RE seed current and demonstrating its parametric dependence on the Ar density (constant across the radius) in Fig. \ref{fig:model1}(a), and on the TQ duration Fig. \ref{fig:model1}(b). Considering dependence on the Ar density, as \Te{} rises generally the RE seed rises at fixed \nAr{} except for in a region where a non-monotonic dependence is found. Considering the experimental data, the Ar density is sufficiently variant across the radius that little can be said other than the RE current rises in both experiment and modeling, and at high \Te{} both model and experiment predict robust RE generation. Considering the TQ duration as the independent axis (Fig. \ref{fig:model1}[b]), the non-monotonicity with \Te{} is removed and clear trends with \Te{} are found in the model. This also illustrates the compounding effect of the experimentally observed shortening of the TQ with \Te{}. Comparing to experiment, a quite good agreement is found, with confidence intervals nearly overlapping with the model prediction curve. This is perhaps expected as the TQ duration provides a rather strong constraint on the RE generation in the hot-tail mechanism. Interestingly, at the highest \Te{}, a wide range of TQ durations yield a similarly high RE seed production.

\begin{figure}
\centering
\includegraphics[width=0.48\textwidth]{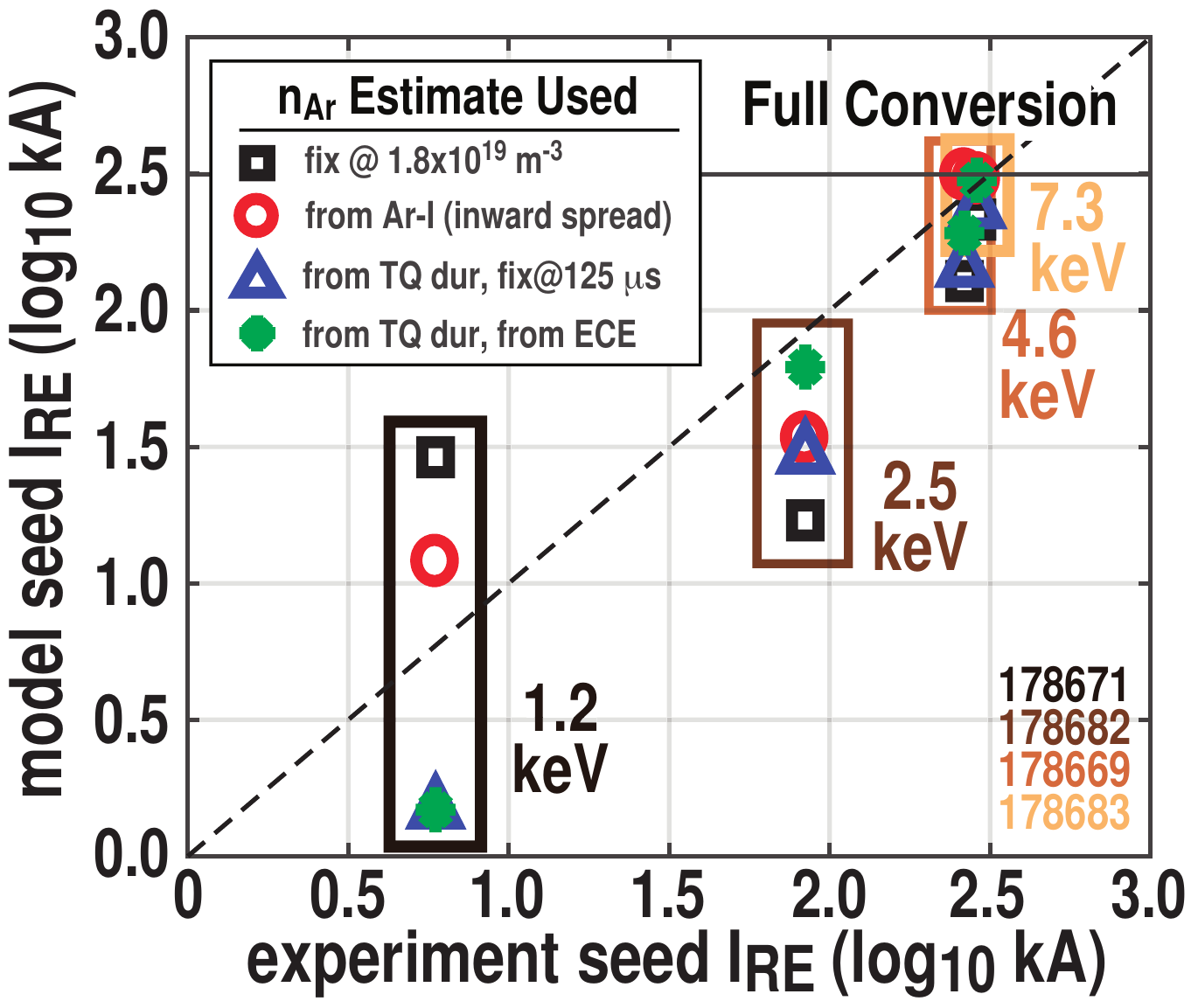}
\caption{Model-experiment comparison of the seed RE current using the various \nAr{} estimates discussed in this section. Model and experiment agree on a robust seed RE production at high \Te{}.}
\label{fig:model2}
\end{figure}

Kinetic modeling of the seed RE current as compared with experimental data using the various \nAr{} estimates (fixed, Ar-I light, TQ duration) of Fig. \ref{fig:modelAr} are systematically compared to experimental values in Fig. \ref{fig:model2}. A clear feature is that at high \Te{} the model predictions using the different estimates converge to a robust RE production, and this is consistent with experiment. At low \Te{}, model estimates are more varying, with the Ar-I emission estimate being the most accurate. Over the full \Te{} range, using the ECE duration is most accurate. A high level conclusion can be made that the experiment traversed from weak to strong RE production, and the kinetic modeling recovers this result.


\section{Full Conversion Regime}
\label{sec:prompt}

Attention now turns to the highest \Teavg{} ($\approx$ 12 keV) discharges that produced REs `naturally' without Ar pellet injection. RE production occurred following a prompt collapse of the ITB together with a significant (but unquantifiable) influx of carbon from the DIII-D first wall. The highest \Teavg{} discharge will be shown to exhibit properties consistent with a prompt conversion of the entire pre-disruption current into REs, resulting in a low energy (sub-MeV) final RE population.

\begin{figure}
\centering
\includegraphics[width=0.4\textwidth]{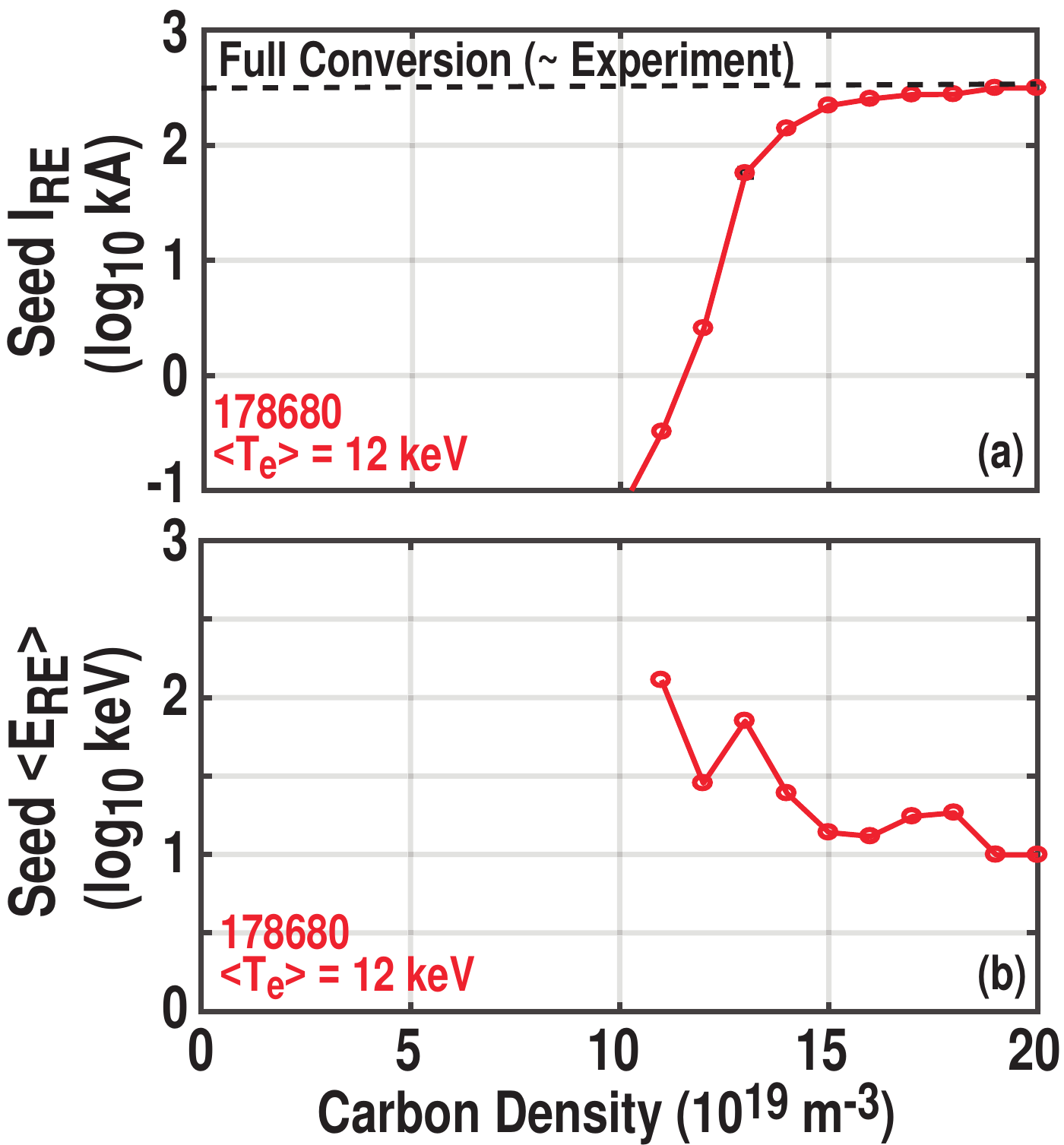}
\vspace{-10 pt}
\caption{Dependence of (a) seed RE current and (b) average seed RE energy on (uniform) carbon density in the highest \Teavg{} discharge accessed.}
\label{fig:modelUFO}
\end{figure}

Kinetic modeling using the same framework described in Sec. \ref{sec:model} is applied to the  \Teavg{} $\approx 12$ keV discharge (red profiles in Fig. \ref{fig:profs}). Note that no radial transport effects are considered in the model. The plasma power balance is still assumed to be governed by impurity line radiation. The dependence of the seed RE current (\Ire{}) on the carbon density is shown in Fig. \ref{fig:modelUFO}, and indicates that at a critical carbon density full conversion of the pre-disruption current to RE current is observed, analogous to argon in Fig. \ref{fig:model1}(a). Fig. \ref{fig:modelUFO}(b) now considers the average seed RE energy \Ereavg{}, indicating that when full conversion of a RE beam occurs a sub-MeV seed \Ereavg{} is expected. Crucially, since the RE conversion is expected to be complete, there is no flux change to further increase the RE energy (no current quench). This peculiar RE population would still carry the current over a cold bulk, but without the MeV-level HXR bremsstrahlung emission characteristic of normal RE beams. On a longer timescale this population is still slowly gaining energy and drifting toward the plateau (near-threshold) regime. But unlike in strongly avalanching cases, \Ereavg{} never exceeds the attractor energy of the plateau \cite{Aleynikov2015}. This regime has previously been observed at low \Ip{} in JET \cite{Reux2017}, though its connection there to \Te{} is unclear.

\begin{figure}
\centering
\includegraphics[width=0.4\textwidth]{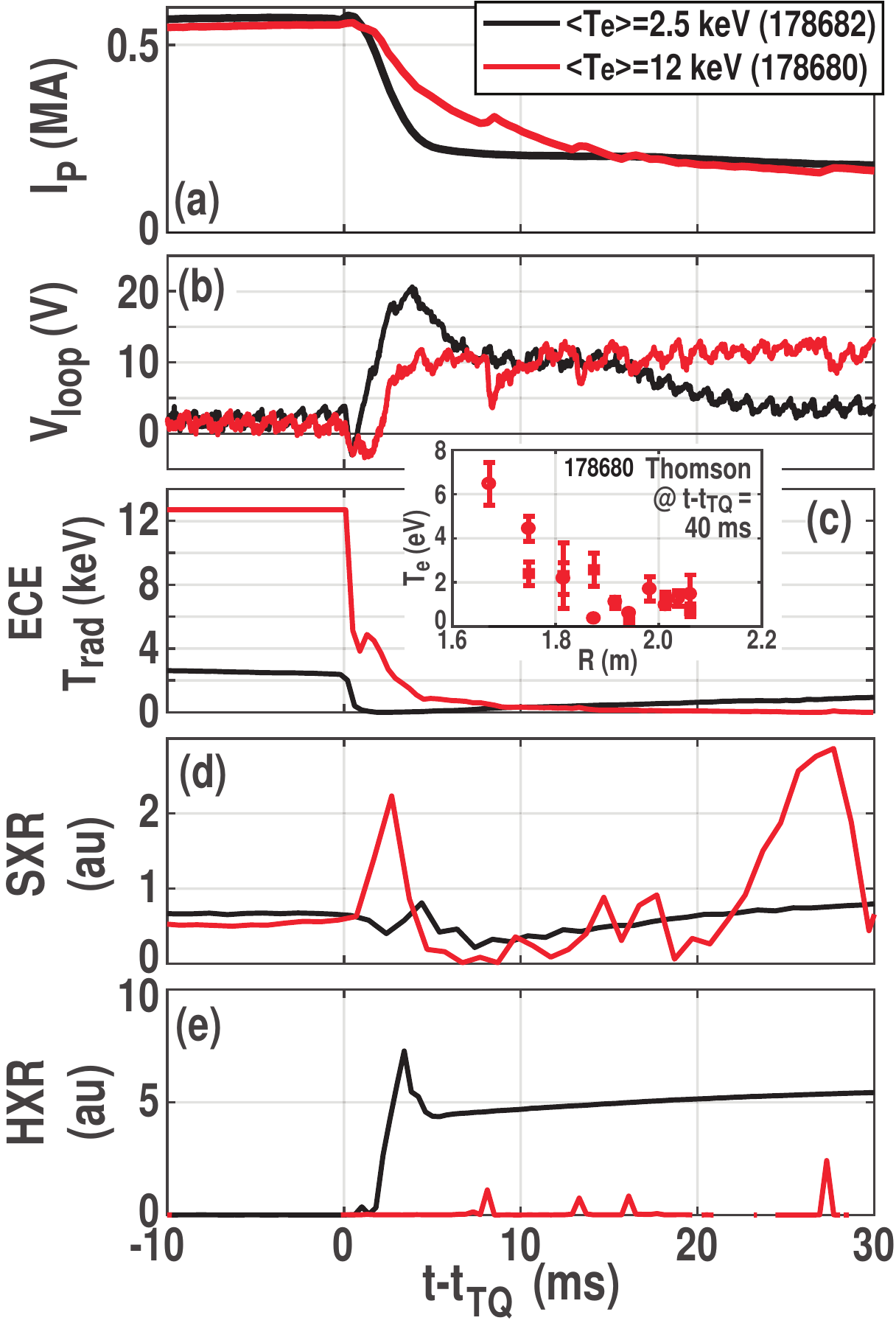}
\vspace{-10 pt}
\caption{Time evolution of discharge undergoing full conversion (red) compared to a lower \Te{} reference (black), indicating (a) plasma current (\Ip{}), (b) loop voltage (\Vloop{}), (c) ECE radiometer \Te{} measurements and with inset post-TQ Thomson scattering \Te{} measurement, (d) soft X-ray emission, and (e) hard X-ray emission.}
\label{fig:UFOtime}
\end{figure}

Experimentally, evidence for sub-MeV RE beams is indeed found in the behavior of the RE beam produced at \Teavg{} of 12 keV. Time-histories are shown in Fig. \ref{fig:UFOtime} comparing a low \Teavg{} discharge (whose shutdown was initiated with an Ar pellet) with the natural RE producing discharge at 12 keV.  The most striking feature of the RE beam initiated from the 12 keV plasma is the absence of significant HXR emission (Fig. \ref{fig:UFOtime}[e]) while SXR emission is still present (Fig. \ref{fig:UFOtime}[d]), strongly supporting a low energy RE beam. Note that the extremely sensitive pulse-height counting Gamma Ray Imager (\cite{Cooper2016,Pace2016,PazSoldan2018,Lvovskiy2018}) does register a very low count rate of HXRs, as even a sub-MeV RE distribution will have a few REs at MeV levels that can emit HXRs.  On DIII-D a RE beam without significant HXR emission is a completely novel observation.

While the conversion to RE current is interpreted to be complete, the \Ip{} still drops after in the high \Teavg{} shutdown, albeit at a much slower rate. This is interpreted to be due to a more resistive than usual final RE beam, with the maximum available loop voltage (\Vloop{}) of 10 V being still below the critical electric field for this RE beam. As such, the final current decreases to the value that can be sustained by \Vloop{}, which is about 0.2 MA. This yields a resistance measure of 50 $\mu \Omega$.  Thomson Scattering provides a measurement of the bulk \Te{} during this phase, finding \Teavg{} $\approx$ 3 eV, and peaking at 6 eV (as shown in the inset of Fig. \ref{fig:UFOtime}[c]). The Spitzer resistivity of the bulk is thus at least an order of magnitude higher than the observed resistivity, confirming that the current must be carried predominantly by REs. Estimating the resistivity of the sub-MeV RE beam formally requires knowledge of the carbon fraction (to estimate the number of bound electrons), which is not known. However, assuming the carbon density to be 19 \densunit{} (near the minimum to access full conversion), and using an appropriate partial screening factor for a sub-MeV population (via Eq. 27 in Ref. \cite{Breizman2019}), the predicted resistivity of the beam is found by the kinetic model to be 60-70 $\mu \Omega$, which is consistent with observation. The better than usual match to experimental resistivity (well predicted collisional dissipation rate) may be due to the absence of high-energy REs and thus an absence of drift-orbit loss, which was recently invoked to explain discrepancies in the observed dissipation rates on DIII-D\cite{Hollmann2019a}. 

The final interesting feature of the full conversion regime is that no modification of the shape control algorithm is in principle needed to maintain control of the RE beam, unlike regular RE beams \cite{Eidietis2012a}. Since the CQ is nominally avoided the conversion from thermal to sub-MeV RE current is transparent to the control system. Indeed no changes were made in the high \Teavg{} discharge shown in Fig. \ref{fig:UFOtime} yet shape control was maintained.


\section{Discussion and Conclusions}
\label{sec:disc}

This study has identified several important features of RE production going from low to high \Te{}, spanning a full reactor-relevant range from 1 to 12 keV. Findings are summarized as follows:

1) The seed RE production is found to be extremely sensitive to \Te{}, and the experiment transitions from modest RE production to prodigious RE production at the highest \Te{}. This is all the more remarkable since the avalanche gain of these plasmas is very small (as discussed in Sec. \ref{sec:seed}). The small avalanche gain fortuitously allows the observed final RE current to be translated to the seed RE current while introducing minimal experimental uncertainty. 

2) The RE production is even more sensitive to \Te{} than expected due to indirect effects that shorten the TQ duration. High \Te{} produces intense pellet ablation that deposits a much larger quantity of radiators into the plasma, leading to a faster TQ (assuming the cooling is determined by line radiation). Alternatively, parallel transport scales like \Te{}$^{7/2}$, leading to another mechanism that could also accelerate the TQ.  If the observed shortening of the TQ with increasing \Te{} extends to fusion-grade plasmas, than their propensity to form large quantities of RE seeds at high \Te{} may be far worse than previously considered.

3) At the highest \Te{} of near 12 keV, observations support the full conversion of the pre-disruption current into sub-MeV RE current without significant HXR emission. This RE beam was naturally kept under shape control (without changing any settings), confirming the control system transparency of full RE conversion and allowing the possibility of a regular soft-landing for these beams. Indeed, a full RE conversion may be preferable than a partial conversion so long as the loss of shape control and vertical instability is avoided.

4) Kinetic modeling robustly reproduces the strong seed RE production at the highest \Te{}, with several estimates of the poorly constrained Ar density profile all leading to strong RE production. Overall, the best agreement with experiment is found when matching the TQ duration, though the measured Ar density assuming radially inward transport also yields agreement within an order of magnitude.

5) \ref{sec:gri} shows the RE energy measured in the early CQ decrease with pre-disruption \Te{}, though the effect cannot be isolated to seed RE dynamics as opposed to differences in the CQ. This can be considered a positive by-product of the increased current conversion to REs.

These observations highlight the propensity to form RE beams in high \Te{} plasmas, which may be even more severe than previously appreciated, and motivates the development of improved theoretical models to better capture the observed effects. While overall this work presents a rather troubling set of findings, a few additional comments can be drawn.

First, this study was dependent on an ITB scenario to reach high \Te{}. This brings the question of whether some observations are peculiar to an ITB scenario as opposed to being generic to any high \Te{} plasma. It is conceivable that a process intrinsic to the ITB (namely a MHD-driven fast crash) is responsible for the observed short TQ durations. It is interesting to note that ITBs were found in JET to have the shortest TQ durations \cite{Riccardo2005}, though they were also among JET's highest \Te{} plasmas (as they are in DIII-D).  The measurements presented in Sec. \ref{sec:MHD} however find no meaningful difference in the observed \dB{} at the wall during the TQ. Furthermore, as discussed in Sec. \ref{sec:pellet}, the injected Ar pellet inventory was more than sufficient to produce the observed TQ durations exclusively through radiation cooling, as presented Sec. \ref{sec:model}. Another possible peculiarity of the ITB is the sharp \Te{} gradient, which would promote intense and local pellet burnup. The sharp \Te{} profile of the ITB scenario however is qualitatively similar to the predicted ITER pedestal \Te{} profile, which is $\approx$ 5 keV at the pedestal top \cite{Baylor2007}. These questions further motivate an integrated treatment of the pellet ablation, RE seed production, and MHD evolution, which will likely require integrating extended MHD codes \cite{Izzo2011,Izzo2012,Bandaru2019,Sommariva2018} with radiation \cite{Lyons2019} and kinetic \cite{Harvey2019} solvers to treat the TQ+RE dynamics.

Secondly, this work points to the importance of doing experimental seed RE studies at high \Te{}. While future reactors will require a disruption mitigation solution that is valid for all \Te{}, high \Te{} plasmas are more relevant to disruptions in a fusion burn scenario and they are found to yield an entirely different dynamic with respect to the pellet ablation and TQ dynamic. The insights gained comparing low and high \Te{} plasma dynamics should be invaluable in developing a believable self-consistent simulation of the RE seed formation process. Varying \Te{} thus offers a fruitful path to validating predictive models in present-day experiments and extrapolating their results to fusion-grade plasmas.

Finally, the high \Te{} plasma presents for the first time a robust platform on DIII-D for demonstrating RE avoidance. Previously, small variations (in \Te{}, or in the pellet integrity) were sufficient to prevent RE formation \cite{James2012,Izzo2012}, making RE avoidance demonstrations meaningless in such marginal conditions. In contrast, a demonstration of RE avoidance via advanced pellet injectors \cite{Hollmann2019,Raman2019} or other techniques \cite{Smith2013,Lvovskiy2019,Guo2018} would be more meaningful in the robustly RE producing high \Teavg{} ITB scenario, though the strong avalanche multiplication would still be absent.

\section*{Acknowledgments}

The authors would like to thank L. Bardoczi, T. Carlstrom, J. Herfindal, L. Stagner, R. Sweeney, and P. Parks for valuable discussions and support in executing this experiment.  

This material is based upon work supported by the U.S. Department of Energy, Office of Science, Office of Fusion Energy Sciences, using the DIII-D National Fusion Facility, a DOE Office of Science user facility, under Awards DE-FC02-04ER54698,DE-AC05-00OR22725, DE-FG02-07ER54917. DIII-D data shown in this paper can be obtained in digital format by following the links at \url{https://fusion.gat.com/global/D3D_DMP}. Disclaimer: This report was prepared as an account of work sponsored by an agency of the United States Government. Neither the United States Government nor any agency thereof, nor any of their employees, makes any warranty, express or implied, or assumes any legal liability or responsibility for the accuracy, completeness, or usefulness of any information, apparatus, product, or process disclosed, or represents that its use would not infringe privately owned rights. Reference herein to any specific commercial product, process, or service by trade name, trademark, manufacturer, or otherwise, does not necessarily constitute or imply its endorsement, recommendation, or favoring by the United States Government or any agency thereof. The views and opinions of authors expressed herein do not necessarily state or reflect those of the United States Government or any agency thereof.


\appendix
\section{RE Energy Spectrum During Early CQ}
\label{sec:gri}

In this Appendix the RE energy measured in the early CQ will be shown to decrease with pre-disruption \Te{}.  Measurements in the early CQ cannot be directly linked to the RE seed because during the CQ, additional dynamics such as flux conversion, RE acceleration, and secondary avalanche take place which are not accounted for in the TQ kinetic modeling of Sec. \ref{sec:model}. Measurements are made with a single central tangentially viewing chord of the Gamma Ray Imager diagnostic \cite{Cooper2016,Pace2016,PazSoldan2018,Lvovskiy2018}, used here functionally as a single HXR spectrometer. HXR spectra can only be measured once the RE energy exceeds 1 MeV, which requires a fraction of the current quench (CQ) to have elapsed. With these limitations in mind, a qualitative comparison to kinetic modeling is nonetheless presented in this Appendix.

\begin{figure}
\centering
\includegraphics[width=0.4\textwidth]{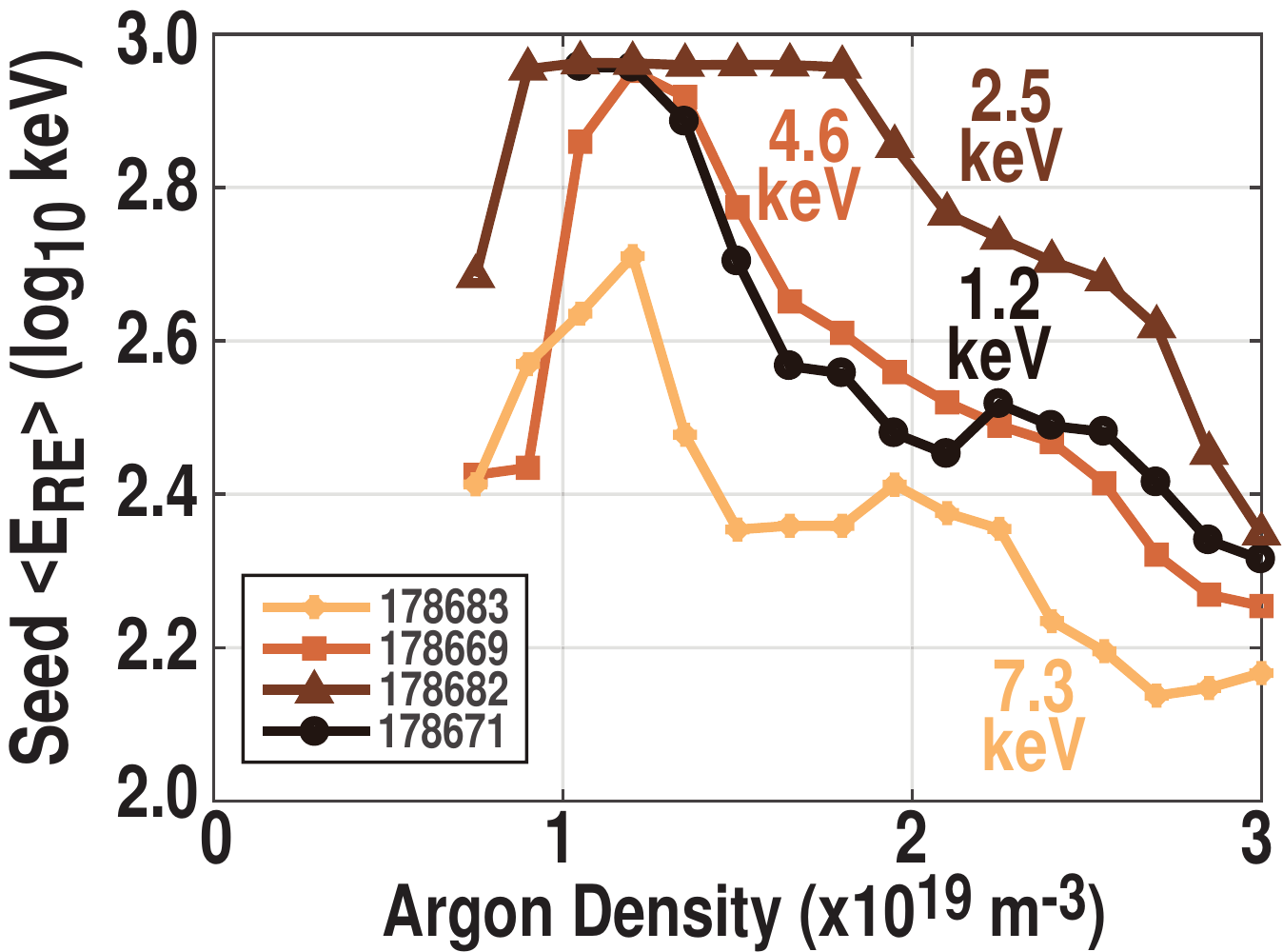}
\caption{Kinetic model prediction of average RE energy (\Ereavg{}) as (uniform) Ar density is increased for the discharges studied in Sec. \ref{sec:model}. As seed Re current increases (Fig. \ref{fig:model1}), \Ereavg{} decreases as more REs are participating in carrying the current.}
\label{fig:EREmodel}
\end{figure}

Kinetic model predictions of the average seed RE energy (\Ereavg{}) are presented in Fig. \ref{fig:EREmodel}, which uses the same abscissa as Fig. \ref{fig:model1}(a). As Ar quantity increases the TQ duration decreases, the RE current conversion increases and \Ereavg{} decreases. This is because more REs are available to carry the current, and so the requisite energy per RE decreases. Furthermore, as \Teavg{} increases, the predicted \Ereavg{} decreases as well, also because more seed REs are available to carry the current. Considering the radial profile, as shown in Fig. \ref{fig:modelAr}, the decreasing \Ereavg{} can also be understood as an expanding spatial region of 100\% conversion, as these regions have lower \Ere{}.

\begin{figure}
\centering
\includegraphics[width=0.4\textwidth]{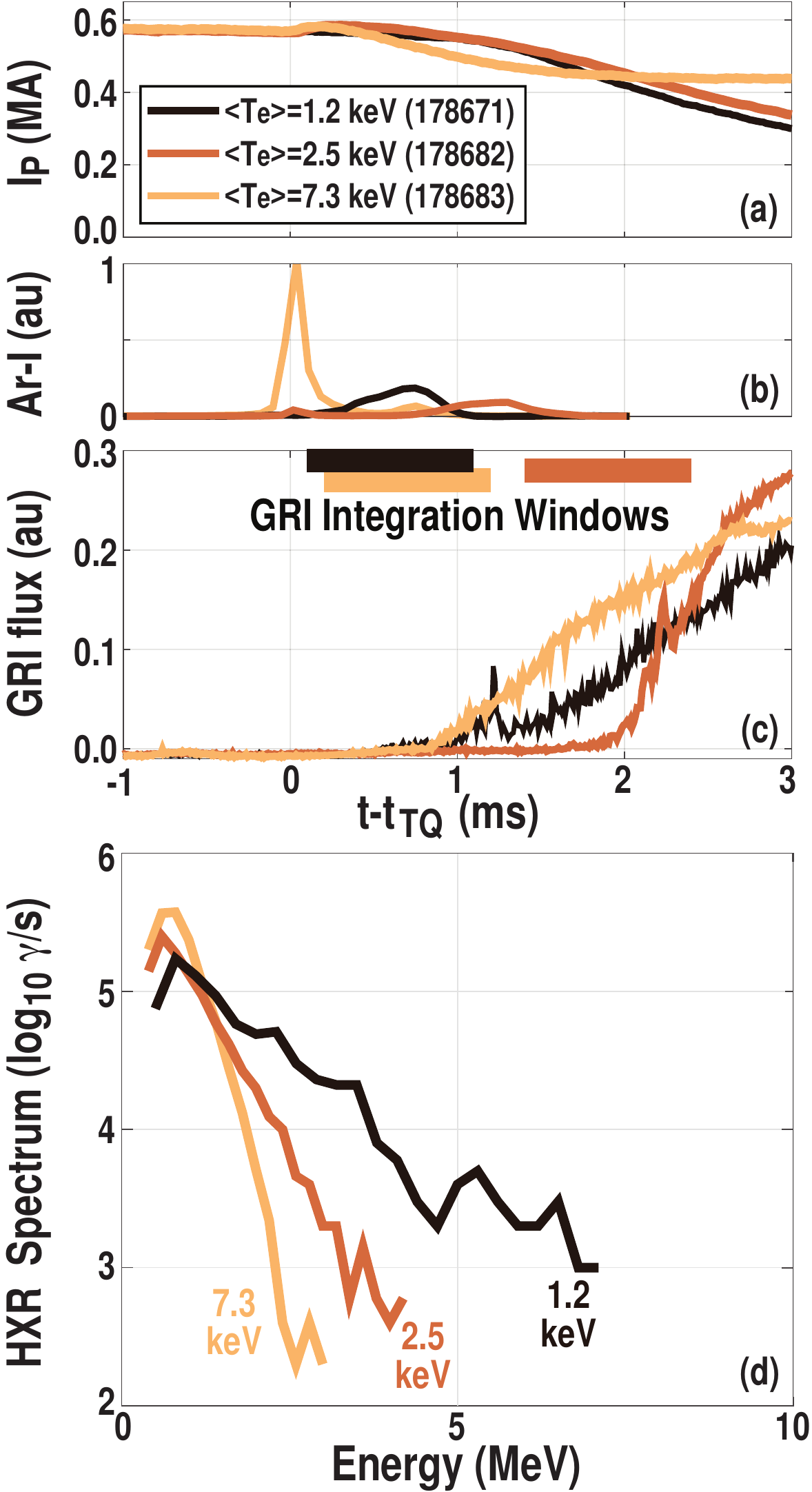}
\caption{GRI measurements for low and high \Te{} discharges, with (a) \Ip, (b) Ar-I ablation light, (c) GRI raw flux, and (d) HXR spectra accumulated via GRI pulse-height counting shown.}
\label{fig:GRI}
\end{figure}

GRI measurements as \Teavg{} is varied are shown in Fig. \ref{fig:GRI}, with high and low \Teavg{} cases highlighted. Note the earlier appearance of HXR flux ($>$ 1 MeV) at high \Te{}, indicating more prompt RE production. Integration windows of 1 ms are used to produce the HXR spectra shown in Fig. \ref{fig:GRI}(c). As can be seen, the HXR spectrum is found to shift to lower energy with \Teavg{}, indicating the RE energy decreases with \Teavg{}. 

While CQ dynamics are outside the scope of this manuscript, it is worth pointing out a few interesting features. Firstly, MHz-range Alfv\'enic instabilities during the CQ (described in Ref. \cite{Lvovskiy2018}) are only observed at the lowest \Te{}, supporting the hypothesis that a minimum \Ereavg{} is required to excite these instabilities. Second, the final post-TQ \Te{} is not constant in the discharges of Fig. \ref{fig:GRI}. For initial \Te{} of 1.2, 2.5, 7.3 keV, the final \Te{} is measured by Thomson scattering to be 3, 5, 2 eV. Understanding the CQ dynamic of these discharges is left to future study.

\begin{figure}
\centering
\includegraphics[width=0.48\textwidth]{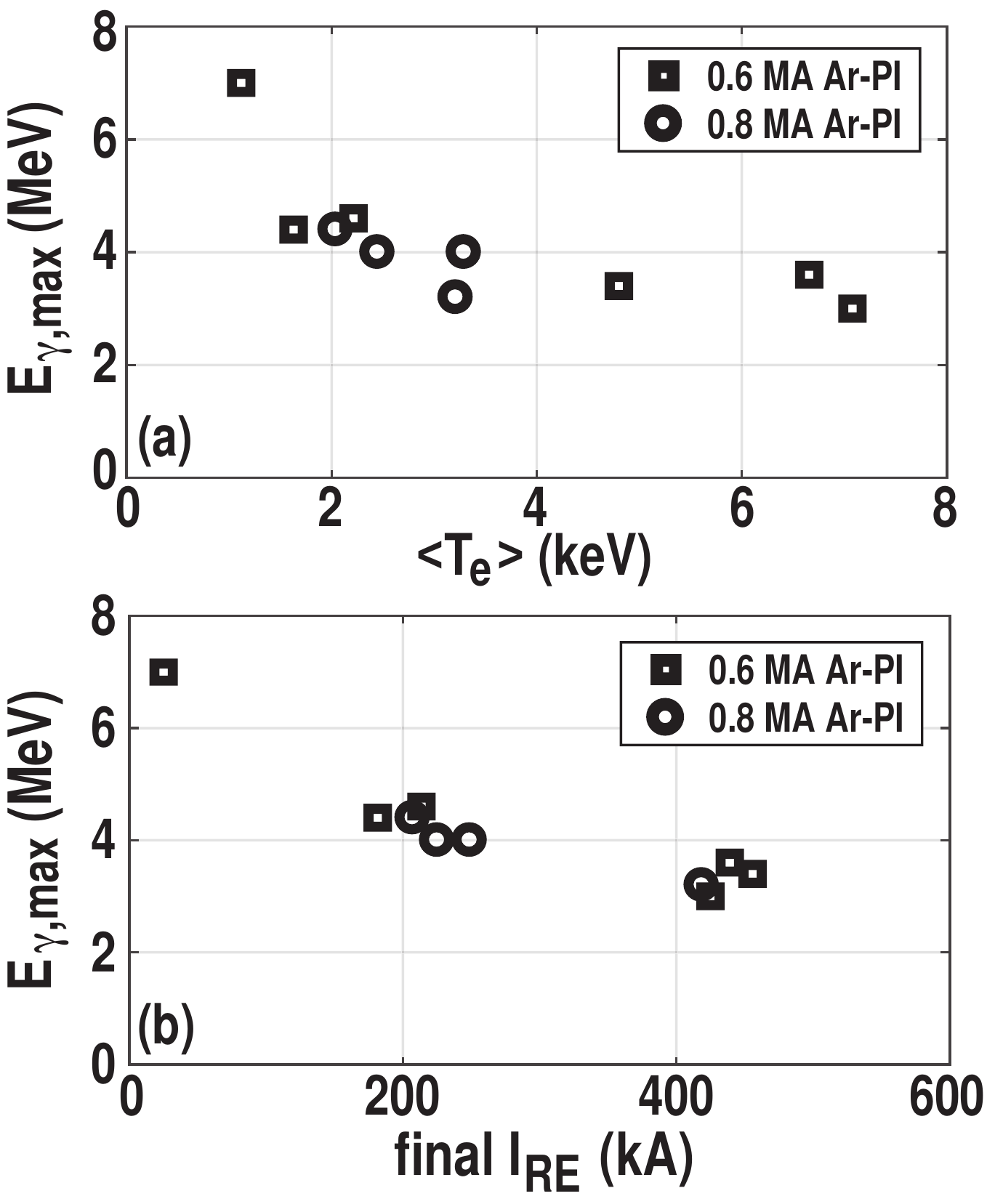}
\caption{Systematic variation of maximum HXR energy (\Egammax{}) vs (a) pre-disruption \Teavg{} and (b) final RE current (\Ire{}).}
\label{fig:GRImetric}
\end{figure}

The maximum HXR energy (\Egammax{}) is plotted against \Te{} and \Ire{} in Fig. \ref{fig:GRImetric}. The first panel can be thought of as a TQ-dominant interpretation, whereby \Egammax{} is set by the seed \Ereavg{}. The second panel is a CQ-dominant interpretation, whereby the more RE current the less flux is available to accelerate the seed REs to higher energy. In reality, it is likely that a combination of both effects is responsible for the decreased RE energy with \Teavg{}.

\bibliographystyle{iopart-num}


\section*{References}

\providecommand{\newblock}{}


\end{document}